\documentclass[a4paper,11pt,oneside,final,openbib]{article}
\pdfoutput=1
\usepackage{graphicx}
\usepackage{amsmath}
\usepackage{amsfonts}
\usepackage{amssymb}
\usepackage{slashed}
\usepackage{bigints}
\usepackage{multirow}
\usepackage{tikz}
\usepackage{float}
\usepackage[margin=1in]{geometry}
\usepackage[toc]{appendix}
\usepackage[utf8]{inputenc}
\usepackage{url}
\usepackage{cite}
\usepackage{hyperref}
\usepackage{breakurl}
\usepackage{physics}

\hypersetup{
    breaklinks=true,
    colorlinks=true,
    linkcolor=blue,
    filecolor=magenta,
    urlcolor=blue,
    citecolor=blue,
}

\newcommand{\li}{\textrm{Li}_{2}}
\def\lsim{\:\raisebox{-0.5ex}{$\stackrel{\textstyle<}{\sim}$}\:}

\begin{document}

\mbox{ }
\hfill MITP/18-108
\\
\mbox{ }
\hfill \today \\

\begin{center}
{\LARGE {\bf 
Second-Order Leptonic Radiative Corrections 
\\[1ex]
for Lepton-Proton Scattering}}
\\
\end{center}
\vspace{.05cm}
\begin{center}
{\bf R.-D. Bucoveanu} $^{(a)}$, 
{\bf H. Spiesberger} $^{(b),(c)}$
\\ 
\end{center}

\begin{center}
{\it $^{(a)}$
PRISMA Cluster of Excellence, 
Institut f\"{u}r Kernphysik, 
Johannes Gutenberg-Universit\"{a}t, D-55099 Mainz, Germany}
\\

{\it $^{(b)}$
PRISMA Cluster of Excellence, 
Institut f\"{u}r Physik, 
Johannes Gutenberg-Universit\"{a}t, D-55099 Mainz, Germany}
\\

{\it $^{(c)}$
Centre for Theoretical and Mathematical Physics, 
and Department of Physics, University of
Cape Town, Rondebosch 7700, South Africa}
\\
\end{center}
\begin{center}
\footnotesize
{\it E-mail:} 
rabucove@uni-mainz.de, 
spiesber@uni-mainz.de 
\end{center}
	
\vspace{16mm}

\begin {abstract}
The interpretation of high-precision lepton-nucleon 
scattering experiments requires the knowledge of 
higher-order radiative corrections. We present a 
calculation of the cross section for unpolarized 
lepton-proton scattering including leptonic radiative 
corrections up to second order, including one- and 
two-loop corrections, radiation of one and two photons 
and one-loop corrections for one-photon radiation. 
Numerical results are given for the planned P2 experiment 
at the MESA facility in Mainz, and some results are 
also discussed for Qweak and the suggested MUSE 
experiment. 
\end{abstract}
\newpage

\section{Introduction}
\label{sec_introduction}

Lepton scattering has been, and continues to be, an 
extremely important experimental technique to study 
properties of matter. Elastic and inelastic electron 
nucleon scattering has allowed us to obtain information about 
form factors and structure functions or, at higher energies, 
parton distribution functions. Particularly interesting 
modern research topics include the investigation of the 
spin structure of the proton or the recently observed 
discrepancy in the determination of the proton charge 
radius between different experimental techniques. 
Precision measurements with polarized electrons are also 
used to study weak interactions. At low energies, with 
elastic electron proton scattering, one can determine the 
weak charge of the proton, which is related to the weak 
mixing angle in the Standard Model. Results of the Qweak 
experiment at the Jefferson Laboratory have been published 
\cite{Androic:2018kni} and the Mainz P2 experiment at the 
MESA accelerator being under construction is expected to 
start commissioning in 2021 \cite{Becker:2018ggl}. Both 
experiments have, or will, also provide new limits for 
physics beyond the Standard Model, complementary to 
searches at high-energy colliders. In addition, a $\mu$p 
scattering experiment, MUSE \cite{Gilman:2013eiv}, has 
been proposed at the PSI with the aim to study the proton 
radius puzzle. 

\bigskip

The improvement of experimental techniques over the years 
has brought high-precision measurements in reach, often 
at the percent level, or even better. It is therefore 
compulsory to improve the calculation of theoretical 
predictions to the same level. Higher-order corrections, 
in particular QED radiative effects, can often not be taken 
from the classical work of Mo and Tsai \cite{Mo:1968cg} 
(see also \cite{Tsai:1961zz,Tsai:1971qi}) without carefully 
revisiting the underlying assumptions and improving 
approximations which had been acceptable in previous 
experiments. Quite a number of articles by different 
groups of authors have appeared since the publication 
of \cite{Mo:1968cg}. Their focus often was put on the 
derivation of more precise explicit and simple to use 
formulas avoiding the soft-photon and peaking 
approximations, see for example \cite{deCalan:1990eb, 
Maximon:2000hm,Akushevich:2015toa,Weissbach:2004ij, 
Weissbach:2008nx}. Also higher-order effects, like 
those due to multi-photon radiation in the soft-photon 
approximation, or re-summed leading logarithms in the 
structure function approach can be found in the literature 
\cite{Arbuzov:2015vba}. 

Radiative corrections depend very strongly on experimental 
details and the way how kinematic variables like energies 
or scattering angles are measured. Therefore calculations 
often require a special treatment for a given experimental 
situation. For example, there are specially crafted 
calculations for ep scattering in coincidence, i.e.\ where 
both the scattered electron and the scattered proton are 
observed \cite{Ent:2001hm}, where only the scattered proton 
is measured \cite{Afanasev:2001nn}, or with reversed 
kinematics, i.e.\ with protons scattering off electrons at 
rest \cite{Gakh:2016xby}. Also the calculation of radiative 
corrections for lepton scattering at very high energies 
requires different techniques \cite{Kwiatkowski:1990es, 
Arbuzov:1995id}.

Radiative corrections for electron scattering can be 
separated into contributions due to (real and virtual) 
photon radiation from the lepton, from the nucleon, and 
its interference. Higher-order effects at the nucleon 
require special attention. Real nucleonic radiation is 
suppressed due to the higher nucleon mass. Apart from its 
role as part of radiative corrections, photon emission from 
the proton is interesting by itself. For large momentum 
transfer it is known as (deeply) virtual Compton scattering 
(DVCS). It is used to study properties of the nucleon, e.g.\ as 
encoded in generalized parton distributions (GPDs). Radiative 
corrections for DVCS involve Feynman diagrams which are 
also part of the second-order radiative corrections studied 
in the present paper, see for example Refs.~\cite{
Vanderhaeghen:2000ws,Bytev:2003qf,Akushevich:2012tw,
Akushevich:2017kct}. 

The interference of radiation from the lepton and from 
the nucleon is intimately linked to two-photon exchange 
graphs (box graphs). Both contributions taken separately 
are infrared divergent, but the infrared divergent terms 
cancel when interference effects and box graphs are combined. 
Two-photon exchange corrections have been scrutinized in 
the recent years, see for example the review in 
Ref.~\cite{Arrington:2011dn}, since they are expected to 
be important when data are analysed with the aim to separate 
the electric and magnetic form factors of the proton. The 
observed discrepancy between different techniques, the 
Rosenbluth separation on the one hand and a technique 
based on polarization measurements on the other hand, 
is sensitive to the treatment of two-photon exchange 
corrections. 

Calculations of these radiative effects connected to the 
nucleon are model-dependent and often depend on 
additional assumptions and approximations (see for example  
\cite{Kuraev:2013dra,Borisyuk:2014ssa,Gerasimov:2016zfr}). 
Soft radiation and virtual effects are, however, not 
observable and appear as a part of the observed, effective 
form factors. The separation of such corrections requires 
a well-defined theoretical definition of bare form factors 
in the first place. Higher-order QED effects at the nucleon 
should be taken into account only if these corresponding 
corrections had been subtracted during data analysis to 
extract the form factors. In practice, the proper inclusion 
of this type of corrections can be a formidable task if the 
treatment of radiative corrections was not well documented 
in publications where form factor parametrizations 
were obtained. The problem is similar to the determination 
of parton distribution functions in high-energy experiments. 
There, a well understood framework based on the factorization 
theorem of perturbative Quantum Chromodynamics and the 
renormalization of parton distribution functions exists; 
however, a systematic approach has not been worked 
out yet for form factor measurements at lower momentum 
transfer. Therefore we do not include a discussion of 
radiative effects from the nucleon in the present paper. 

In a realistic experiment one has to impose a set of 
conditions which fix the observable part of the final-state 
phase space. For example, the scattering angle will be 
restricted by the acceptance of the detector, or the 
energy of final-state particles is limited. If the goal 
is to measure elastic form factors, one will try to reduce 
the impact of non-elastic processes, for example by 
imposing a cut-off on the missing energy. This would remove 
e.g.\ pion production, but also restrict the emission of 
hard photons. In experiments with very high luminosity 
like P2, it is impossible to realize cuts on individual 
scattering events and the feasibility to impose kinematic 
conditions may be restricted. Finally, the efficiency for 
the detection of a scattering event may depend on energies 
and scattering angles and vary considerably over the 
observed phase space. It is therefore obvious that a 
Monte Carlo simulation program of the process, ideally 
interfaced to the simulation code of the detector response, 
is indispensable. This approach has become the standard 
for deep-inelastic lepton scattering like at HERA 
\cite{Kwiatkowski:1990es,Charchula:1994kf}, but has also been 
discussed for elastic ep scattering \cite{Akushevich:2011zy}.
In addition, with nowadays computer resources, computer 
algebra systems and high-performance computing on multi-core 
systems, there is anyway no need anymore to search for 
simplified, i.e.\ approximate expressions which are fast 
to evaluate.

In this paper we re-derive the first-order radiative 
corrections for elastic lepton-nucleon scattering. The 
emphasis is, however, on the description of second-order 
corrections, i.e.\ two-loop and two-photon bremsstrahlung 
for unpolarized lepton proton elastic scattering. As 
explained above, we restrict ourselves to purely leptonic 
corrections, i.e.\ not including 2- or 3-photon 
exchange (box graphs) and not including radiation 
from the proton. The corrections are implemented in a 
new Monte Carlo simulation program for numerical 
calculations which we plan to make publicly available 
in the near future \cite{anaximandros}. Also an extension 
to include QED corrections for the scattering of polarized 
leptons is in preparation. 

The first-order corrections are treated in 
Sec.~\ref{sec_first_order}. In Sec.~\ref{sec_second_order} 
we describe our new calculation of second-order 
corrections, including non-radiative parts and 
corrections due to the radiation of one or two 
photons. Then we describe some tests of the implementation 
in a program package for numerical evaluations in 
Sec.~\ref{sec_tests}. Finally, in Sec.~\ref{sec_results} 
we present some numerical results, first of all 
for applications at the forthcoming P2 experiment in 
Mainz, and we conclude with final remarks in 
Sec.~\ref{sec_conclusion}.

\section{Definitions and general remarks}
\label{sec_basics}

\begin{figure}[b]\centering
\includegraphics[width=0.45\textwidth]{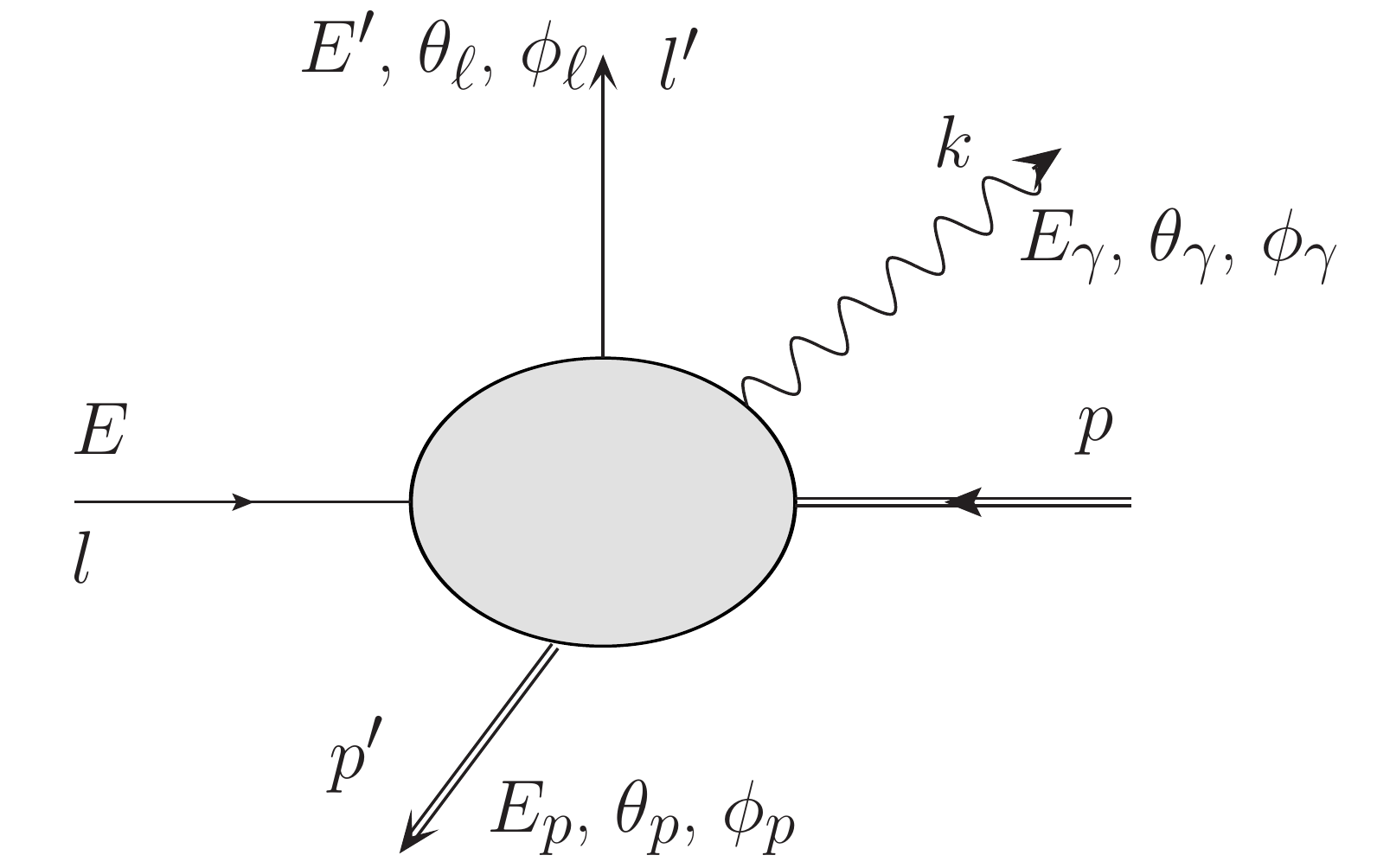}
\caption{Definition of kinematic variables for lepton 
nucleon scattering.
} 
\label{Fig:process_kin}
\end{figure}

We denote the 4-momenta of the incoming and scattered 
lepton (nucleon) by $l^\mu$ and $l^{\prime\,\mu}$  ($p^\mu$ 
and $p^{\prime\,\mu}$). According to the applications 
considered in this work we choose a coordinate frame where 
the target nucleon is at rest and the $z$ axis is directed 
along the momentum of the incident lepton. Symbols for 
energies and angles of the particles involved in the 
scattering process can be found in Fig.~\ref{Fig:process_kin}.

At lowest order, lepton nucleon scattering is described by the 
exchange of a virtual photon. The spin-averaged matrix element 
of the electromagnetic current for a spin-1/2 nucleon, 
$e\bar{u}(p^\prime) \Gamma^\mu_p u(p)$, can be decomposed into 
Dirac and Pauli form factors, $F_1^p$ and $F_2^p$, which are 
functions of the momentum transfer $Q^2 = - (p - p^\prime)^2$. 
They are normalized by $F_1^p(0) = 1$ and $F_1^p(0) = \kappa_p 
- 1$ with $\kappa_p = 2.7928$ the proton's magnetic moment 
in units of the nuclear magneton. The proton vertex is determined 
by 
\begin{equation}
\label{eq_proton_vertex}
\Gamma^\mu_p 
= \gamma^\mu (F_1^p(Q^2) + F_2^p(Q^2)) 
- \frac{(p + p^\prime)^\mu}{2M} F_2^p(Q^2) 
\end{equation}
where $M$ is the nucleon mass. From this vertex rule one 
obtains the tree-level cross section for the scattering of 
unpolarized leptons  with mass $m_\ell$ off unpolarized 
nucleons: 
\begin{eqnarray}
\frac{\mathrm{d}\sigma^{(0)}}{\mathrm{d}Q^2} 
&=& 
\frac{\pi\alpha^2}{Q^4 
\left((s - m_\ell^2 - M^2)^2 - m_\ell^2 M^2 \right)} 
\\ & &  
\Bigl[
2 \left(F_1^p\right)^2 
\left((s - m_\ell^2 - M^2)^2 + (s-Q^2)^2 - 2 s (m_\ell^2 + M^2) 
      + (M^2 + m_\ell^2)^2 \right) 
\nonumber \\ & &  
+ 4 F_1^p F_2^p Q^2 \left(Q^2 - 2 m_\ell^2 \right)
\nonumber \\ & &   
+ \left(F_2^p\right)^2 \frac{Q^2}{M^2} 
\left((s - m_\ell^2 - M^2)(s - Q^2) - (s-Q^2) M^2  
- s m_\ell^2 + (M^2 - m_\ell^2)^2 \right)
\Bigr] \, .
\nonumber
\end{eqnarray}
(see \cite{Drell:1963ej}). 
Here, $s$ is the square of the energy in the center-of-mass 
reference frame, $s = (l + p)^2$ and $\alpha$ the fine 
structure constant. An alternative compact expression for 
the Born cross section including all mass terms is given 
in App.~\ref{app_me}. 

For electron scattering it is often a very good approximation 
to neglect the lepton mass. For the applications we have in 
mind, the energies of the incoming and scattered electrons, $E$ 
and $E^\prime$, are large, $E, E^\prime \gg m_\ell$. For not 
too small lepton scattering angles $\theta_\ell$ one can also 
assume that $m_\ell^2 \ll Q^2$. It is also often convenient 
to use the Sachs electric and magnetic form factors $G_E = 
F_1^p - \tau F_2^p$ and $G_M = F_1^p + F_2^p$ with $\tau = 
Q^2/(4M^2)$ and the cross section can be written in the more 
compact form, neglecting terms with the lepton mass,  
\begin{equation}
\frac{\mathrm{d}\sigma^{(0)}}{\mathrm{d}\Omega_\ell} 
= 
\frac{\mathrm{d}\sigma_{\text{Mott}}}{\mathrm{d}\Omega_\ell} 
\frac{1}{\epsilon (1 + \tau)} 
\left[
\epsilon G_E^2(Q^2) + \tau G_M^2(Q^2)
\right]
\end{equation}
with 
\begin{equation}
\frac{\mathrm{d}\sigma_{\text{Mott}}}{\mathrm{d}\Omega_\ell} 
= 
\frac{\alpha^2}{4E^2} 
\frac{\cos^2(\theta_\ell/2)}{\sin^4(\theta_\ell/2)}
\frac{E^\prime}{E} 
\end{equation}
and 
\begin{equation}
\epsilon = 
\left[
1 + 2(1+\tau) \tan^2(\theta_\ell/2)
\right]^{-1} 
\, .
\end{equation}
We note that our program for the numerical evaluation of 
cross sections includes all lepton mass terms and is applicable 
also for the case of muon scattering. 

The proton form factors are considered as external input 
and have to be extracted from measurements. We have 
implemented 5 different types of parametrizations existing 
in the literature. This can be easily modified, if needed. 
All our results shown below are obtained with a simple 
dipole form factor parametrization, $G_E = \left(1 + Q^2 / 
\Lambda^2\right)^{-2}$ and $G_M = \kappa_p G_E$ with 
$\Lambda = 0.71$~GeV$^2$. 

At the tree-level and without the emission of a hard photon, 
the momentum transfer to the nucleon can be determined from 
the energy and scattering angle of the outgoing lepton. 
However, bremsstrahlung can lead to a shift of $Q^2$ and we 
have to distinguish the value determined from the scattered 
lepton from its true value transferred to the nucleon. To 
emphasize this fact, we use the additional symbol $Q_\ell^2$, 
defined by
\begin{equation}
Q_\ell^2 = - (l - l^\prime)^2 \, .
\end{equation}

\bigskip

Higher-order corrections are due to additional photon emission 
and absorption, either virtual, described by loop diagrams, or 
real, described by bremsstrahlung diagrams. Both parts contain 
infrared (IR) divergences which cancel when combined. In our 
approach we use the phase-space slicing method to separate 
soft-photon radiation from hard-photon contributions. 
The separation is implemented by using a cut-off $\Delta$ 
for the energy of a radiated photon. $\Delta$ is chosen 
small, below the detection threshold for the observation 
of a photon in the detector, and the soft-photon part 
combined with loop diagrams is called non-radiative. 
First-order corrections, at order $\mathcal{O}(\alpha)$ 
relative to the Born cross section, are written as 
\begin{equation}
\sigma^{(1)} 
= \sigma^{(1)}_{\text{non-rad}} + \sigma^{(1)}_{1h\gamma} \, ,
\end{equation}
and 
\begin{equation}
\sigma^{(1)}_{\text{non-rad}} 
= \sigma^{(1)}_{1-\text{loop}} + \sigma^{(1)}_{1s\gamma} \, .
\end{equation}
At second relative order, one has to include contributions 
with both one or two radiated photons and one has to 
distinguish the cases where only one or both photons 
are either soft or hard. The second-order contribution to 
the cross section, $\sigma^{(2)}$ is therefore split into 
three parts:   
\begin{equation}
\sigma^{(2)} 
= \sigma^{(2)}_{\text{non-rad}} 
+ \sigma^{(2)}_{1h\gamma} 
+ \sigma^{(2)}_{2h\gamma},
\end{equation}
where 
\begin{equation}
\sigma^{(2)}_{\text{non-rad}} 
= \sigma^{(2)}_{2-\text{loop}} 
+ \sigma^{(2)}_{1-\text{loop} + 1s\gamma} 
+ \sigma^{(2)}_{2s\gamma}
\, , \quad \quad 
\sigma^{(2)}_{1h\gamma} 
= \sigma^{(2)}_{1-\text{loop} + 1h\gamma} 
+ \sigma^{(2)}_{1s\gamma + 1h\gamma} \, .
\end{equation}
The non-radiative parts are rendered IR-finite by including 
loop diagrams: $\sigma^{(2)}_{\text{non-rad}}$ contains two-loop 
contributions and mixed soft-photon + one-loop parts, while 
$\sigma^{(2)}_{1h\gamma}$ contains one-loop corrections to the 
radiative process with one hard photon. 

In addition we will use correction factors defined relative 
to the differential Born-level cross section 
$\mathrm{d}\sigma^{(0)}$, 
\begin{equation}
\sigma^{(1)}_{\text{non-rad}} 
= \int \mathrm{d}\sigma^{(0)} 
\Big[ \delta^{(1)}_{1-\text{loop}} + 
\delta^{(1)}_{1s\gamma}(\Delta) \Big],
\end{equation}
\begin{equation}
\sigma^{(2)}_{\text{non-rad}} 
= \int \mathrm{d}\sigma^{(0)} 
\Big[ \delta^{(2)}_{2-\text{loop}} 
+ \delta^{(2)}_{1-\text{loop}+1s\gamma}(\Delta) 
+ \delta^{(2)}_{2s\gamma}(\Delta) \Big],
\end{equation} 
where each $\delta$ is labeled with indices as described 
above for the total cross sections. We also show explicitly the 
dependence of the soft-photon parts on the IR cut-off $\Delta$. 
The soft-photon part can be calculated analytically, integrating 
up to the cut-off $\Delta$, by using a soft-photon 
approximation as described in Sec.~\ref{sec_2sgamma_cs} 

Contributions with a hard photon, i.e.\ with energy above 
the cut-off $\Delta$, are infrared finite and the phase 
space integration can be performed numerically. For one 
hard photon at tree level, we can write
\begin{equation}
\sigma^{(1)}_{1h\gamma} = \int _{E_\gamma > \Delta} 
\mathrm{d}^4\sigma^{(1)}_{1\gamma} 
\, ,
\end{equation}
while at second order we define relative correction factors 
for the one-loop and soft photon contributions by writing 
\begin{equation}
\sigma^{(2)}_{1h\gamma} = \int _{E_\gamma > \Delta} 
\mathrm{d}^4\sigma^{(1)}_{1\gamma} 
\left[\delta^{(1)}_{1-\text{loop}+1h\gamma} 
+ \delta^{(1)}_{1s\gamma+1h\gamma}(\Delta) \right] \, , 
\label{eq:sigma2-1gh}
\end{equation}
where $\mathrm{d}^4\sigma^{(1)}_{1\gamma}$ is the differential 
cross section for one radiated hard photon at the tree-level. 
The calculation of $\sigma^{(2)}_{1h\gamma}$ is treated in 
Sec.~\ref{sec_1sgamma_1hgamma_cs}. Finally, the cross section 
for two hard photons is given by
\begin{equation}
\sigma^{(2)}_{2h\gamma} = \int_{E_\gamma , \; E'_\gamma > \Delta} 
\mathrm{d}^7\sigma^{(2)}_{2\gamma} \, .
\end{equation}
It is free of any infra-red singularities and can be calculated 
numerically as described in section \ref{sec_2hgamma_cs}. We 
emphasize that the cut-off parameter $\Delta$ is introduced 
only for a technical reason: it allows us to separate the 
IR singularities. Only separate parts contributing to the 
cross section carry a $\Delta$-dependence as shown in the 
formulas given above. The sum of non-radiative and hard-photon 
contributions has to be independent of $\Delta$. However, 
when we use the soft-photon approximation to calculate the 
non-radiative contributions, and due to numerical uncertainties 
we don't expect the result to be exactly $\Delta$-independent. 
We will study this at more detail below. 

\bigskip

Explicit simple expressions for Feynman diagrams with loops 
can be found in the literature. Where necessary, we use the 
Mathematica package Feyncalc \cite{Shtabovenko:2016sxi} to 
perform the calculations, including a reduction to the 
conventional scalar one-loop Passarino-Veltman integrals 
$B_0$, $C_0$ and $D_0$. The final result is obtained in 
terms of scalar integrals and kinematic invariants. Where 
possible we use explicit simple expressions for the scalar 
integrals. For the numerical evaluation of more complex scalar 
integrals we use the package LoopTools \cite{Hahn:1998yk}.

\section{First-order corrections}
\label{sec_first_order}

\subsection{Non-radiative cross section}

\subsubsection{One-loop corrections}

\begin{figure}[b]\centering
  \includegraphics[width=0.6\textwidth]{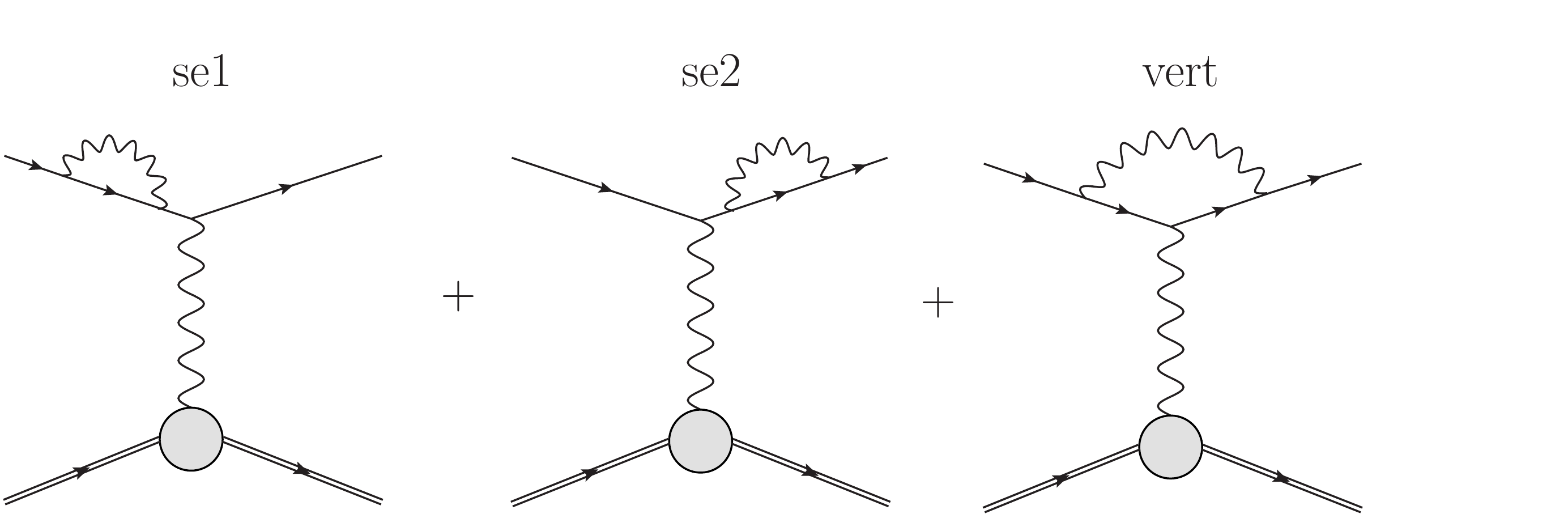}
  \caption{
  Feynman diagrams for the one-loop corrections at 
  the lepton line.
  }
  \label{1st_order_photonic_unpol}
\end{figure}

The one-loop corrections at the lepton line include self energy 
diagrams at the external lines and the vertex graph. The 
first-order correction to the matrix element squared is given 
by 
\begin{equation}
\left|\mathcal{M}_\text{Born}\right|^2	
+ 2\Re(\mathcal{M}_\text{Born}^\dagger \mathcal{M}_\text{se1+se2})
+ 2\Re(\mathcal{M}_\text{Born}^\dagger \mathcal{M}_\text{vert}) 
\end{equation}
where the meaning of the labels for the separate contributions 
corresponds to the diagrams shown in 
Fig.~\ref{1st_order_photonic_unpol}. 
The free lepton propagator for a lepton with four-momentum 
$l$, $S^{(0)}(l)$ is modified by the self-energy $\Sigma (l)$, 
\begin{equation}
S(l)=S^{(0)}(l) + S^{(0)}(l)\Sigma(l)S(l).
\end{equation}
The renormalized self-energy is given by 
\begin{equation}
\Sigma^R (l) 
= \Sigma (l) - (\slashed{l} - m_\ell) \delta Z_2 - \delta m_\ell,
\end{equation}  
where the counter-terms $\delta Z_2$ and $\delta m_\ell$ contain 
the ultra-violet (UV) divergences and are given, in dimensional 
regularisation, by
\begin{align}
\label{eq_delta_Z2}
\delta Z_2 
&= 
- \frac{\alpha}{4\pi} 
\left(\Delta_\epsilon - \log \frac{m_\ell^2}{\mu^2} 
+ 2 \log \frac{\lambda^2}{m_\ell^2} + 4 \right),  
\\
\delta m_\ell 
&= 
\frac{\alpha}{4 \pi} m_\ell 
\left(3 \Delta _\epsilon - 3 \log \frac{m_\ell^2}{\mu ^2} + 4\right). 
\end{align}
$\Delta_\epsilon = \frac{2}{\epsilon} - \gamma_E + \log 4 \pi$ 
contains the $1/\epsilon$-poles of the UV divergences, $\mu$ 
is the mass scale parameter of dimensional regularization and 
$\lambda$ is a finite photon-mass used to regularize the IR 
divergence. For on-shell leptons, the self-energy diagrams 
vanish after renormalization and we only need to include the 
vertex diagram. The relative one-loop correction is given by
\begin{equation}\label{eq_delta_1loop}
\delta^{(1)}_{1-\text{loop}} 
= 
\frac{2\Re(\mathcal{M}_\text{Born}^\dagger 
\mathcal{M}_\text{vert})}{\left| \mathcal{M}_\text{Born} \right|^2}.
\end{equation}
For on-shell leptons, the vertex correction can be separated 
into two form factors, similar to the case of the photon-nucleon 
form factors. Therefore the correction can be taken into account 
by replacing the tree-level on-shell vertex by 
\begin{equation}
\gamma^\mu \rightarrow \Gamma^\mu_\text{vert} 
\equiv 
F^\ell_1 (Q_\ell^2) \gamma^\mu 
+ \frac{i}{2 m_\ell} \sigma_{\mu \nu} q^\nu F^\ell_2 (Q_\ell^2) 
\, .
\end{equation}
$F^\ell_2$ is UV and IR finite. It is proportional to $m_\ell$ 
and therefore very small, but we include it for completeness 
in our calculations. $F^\ell_1$ is both UV and IR divergent. 
The UV divergence is regularized using dimensional 
regularization and removed by renormalization. At first order 
the renormalized form factor is given by 
\begin{equation}
{F^\ell_1}^{(1,R)}(Q_\ell^2) 
= 
{F^\ell_1}^{(1)}(Q_\ell^2) - {F^\ell_1}^{(1)}(0)
= 
{F^\ell_1}^{(1)}(Q_\ell^2) + \delta Z_1 \, ,
\end{equation}   
where we introduced additional upper indices to display 
the loop-order and distinguish renormalized (with index $R$) 
from unrenormalized quantities. The counter-term $\delta Z_1$ 
is given in the $\overline{\text{MS}}$ prescription by
\begin{equation}
\delta Z_1 
= 
- \frac{\alpha}{4\pi} 
\left(\Delta_\epsilon - \log \frac{m_\ell^2}{\mu ^2} 
+ 2 \log \frac{\lambda^2}{m_\ell^2} + 4 \right) \, .
\label{Eq:deltaZ1}
\end{equation}
$\delta Z_1$ is identical with $\delta Z_2$, 
Eq.~\eqref{eq_delta_Z2}, as a consequence of the Ward identity.
The relative vertex correction is therefore given by, up to 
terms suppressed by the lepton mass, 
\begin{equation}
\delta^{(1)}_{1-\text{loop}} = 2 {F^\ell_1}^{(1,R)} \, .
\end{equation}
The result of the loop integration is well-known and can be 
obtained including the exact lepton mass dependence:  
\begin{align}
\begin{split}
\delta^{(1)}_{1-\text{loop}} 
&= 
\frac{\alpha}{\pi} 
\Bigg[ \frac{v^2 +1}{4v}\ln \left(\frac{v + 1}{v -1}\right)
\ln\left(\frac{v^2 -1}{4v^2}\right) 
+ \frac{2v^2 + 1}{2v} \ln \left( \frac{v + 1}{v -1} \right) 
- 2 
\\ 
&+ 
\frac{v^2 + 1}{2v} \left(\li\left( \frac{v+1}{2v}\right) 
- \li \left( \frac{v - 1}{2v}\right)\right) \Bigg] 
+ \delta_\text{IR},
\end{split}
\label{eq_delta_vert}
\end{align}
where $v = \sqrt{1+4m_\ell^2/Q_\ell^2}$ and $\delta_\text{IR}$ 
is the term that contains the IR divergence, given by
\begin{equation}
\label{eq_d_IR}
\delta_{\text{IR}} 
= 
\frac{\alpha}{\pi}\ln(\frac{\lambda^2}
{m_\ell^2})\left[\frac{v^2+1}{2v}\ln(\frac{v+1}{v-1})-1\right] 
\, . 
\end{equation}
This term will cancel at the level of the cross section when 
one-photon radiation is included, as will be seen below. 

Although in most cases it is safe to ignore the $F^\ell_2$ 
form factor, we include it in our calculation, since it 
might become important in some regions of the phase space 
or for the case of $\mu$ scattering. The expression for this 
form factor can be found in \cite{Vanderhaeghen:2000ws} and 
is given, at first order, by
\begin{equation}
{F^\ell_2}^{(1)} = \frac{\alpha}{4 \pi}\frac{v^2-1}{v}
\ln(\frac{v + 1}{v - 1}) \, .
\end{equation}

\subsubsection{One radiated photon in the soft-photon approximation}

\begin{figure}[t]\centering
	\includegraphics[width=0.45\textwidth]{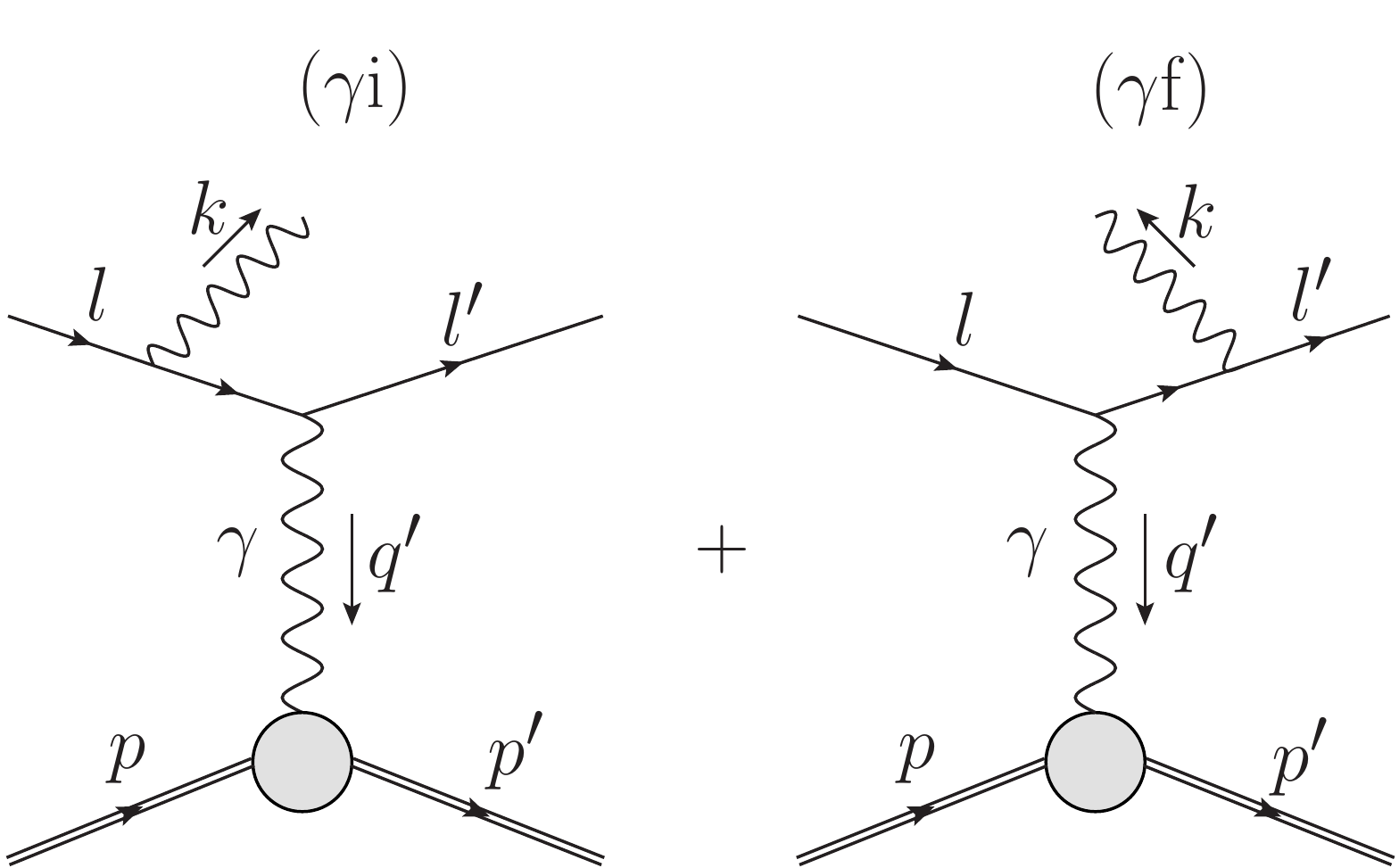}
	\caption{Feynman diagrams for first-order bremsstrahlung 
	corrections.}
	\label{diag_1g}
\end{figure}

The diagrams that contribute to the radiative process $\ell 
p \rightarrow \ell p\gamma$ are shown in Fig.~\ref{diag_1g}. 
We indicate by labels i and f when the photon is emitted 
from the initial or the final lepton. The 4-momentum of 
the additional photon is denoted by $k^\mu$ and its energy 
by $E_\gamma$. The matrix element for radiative scattering is 
\begin{equation}
\label{eq_1g_melem}
\mathcal{M}_{1\gamma} 
= 
\mathcal{M}_{1\gamma\text{i}} + 
\mathcal{M}_{1\gamma\text{f}} \, . 
\end{equation}
Note that the momentum transfer is shifted by the emission 
of a photon, $q^{\mu} \rightarrow q^{\prime \mu} = l^\mu - 
l^{\prime \mu} - k^\mu$, and the matrix element is proportional 
to $1/Q^2$, not $1/Q^2_\ell$. In the soft-photon approximation, 
the matrix element reduces to
\begin{equation}
\label{eq_1sg_melem}
\mathcal{M}_{1\gamma} 
\rightarrow
\mathcal{M}_{1s\gamma} 
= 
- e \mathcal{M}_{0\gamma} 
\left(\frac{l \epsilon^*}{lk} - \frac{l^\prime\epsilon^*}{l^\prime k} 
\right) 
\end{equation}
where $\epsilon^\mu$ is the photon polarization vector. 
Integration over the photon 4-momentum up to a cut-off 
$\Delta$ in the soft-photon approximation leads to
\begin{equation}
\label{eq_cs_soft_photon}
\int_{E_\gamma < \Delta} \mathrm{d}^4\sigma_{1s\gamma}^{(1)}
= 
\mathrm{d}\sigma^{(0)} 
\left(-\frac{\alpha}{2\pi^2}\right)
\int_{E_\gamma < \Delta} \frac{\mathrm{d}^3k}{2 E_\gamma}
\left(\frac{l}{lk} - \frac{l^\prime}{l^\prime k}\right)^2.
\end{equation}
Then the relative one soft-photon correction is given by
\begin{equation}
\label{eq_delta_1sg}
\delta^{(1)}_{1s\gamma}(\Delta) 
= 
\left(-\frac{\alpha}{2\pi^2}\right)
\int_{E_\gamma < \Delta} \frac{\mathrm{d}^3k}{2E_\gamma}
\left(\frac{l}{lk} - \frac{l^\prime}{l^\prime k}\right)^2 
= 
- \frac{\alpha}{\pi} 
\left( B_{ll} - B_{ll^\prime} + B_{l^\prime l^\prime} \right)
- \delta_{\text{IR}} \, ,
\end{equation}
where the result has been written as a contribution from 
initial state radiation, $B_{ll}$, final state radiation, 
$B_{l^\prime l^\prime}$ and the interference between the 
two, $B_{ll^\prime}$. The IR divergence is contained in 
$\delta_{\text{IR}}$ and cancels exactly against the IR 
divergent part of Eq.~(\ref{eq_delta_vert}).
The calculation of $B_{ll}$ and $B_{l^\prime l^\prime}$ is 
straightforward and leads to 
\begin{align}
B_{ll} 
&= 
\ln \left( \frac{2 \Delta}{m_\ell} \right) 
+ 
\frac{E}{|\vec{l}|} 
\ln \left( \frac{m_\ell}{E + |\vec{l}|} \right) \, , 
\\
B_{l^\prime l^\prime} 
&= 
\ln \left( \frac{2 \Delta}{m_\ell} \right) 
+ 
\frac{E'}{|\vec{l}^\prime|} 
\ln \left( \frac{m_\ell}{E' + |\vec{l}^\prime|} \right) \, .
\end{align}
The calculation of the interference term $B_{ll^\prime}$ is 
more involved and can be done following Refs.~\cite{tHooft:1978jhc} 
and \cite{Maximon:2000hm}. The final result is given by 
\begin{align}
\nonumber
B_{ll^\prime} 
&= 
\ln( \frac{4 \Delta^2}{m_\ell^2}) 
\frac{v^2 +1}{v}\ln(\frac{v + 1}{v -1})
+ 
\frac{ \beta ll^\prime}{\xi(\beta E-E')} 
\Bigg[ \ln^2 \left(\frac{E-|\vec{l}|}{E+|\vec{l}|} \right) 
- 
\ln^2 \left(\frac{E'-|\vec{l}^\prime|}{E'+|\vec{l^\prime}|} \right) 
\\
& 
+ \li \left( 1-\frac{\beta(E-|\vec{l}|)}{\xi} \right) 
- \li \left( 1-\frac{\beta(E'-|\vec{l}^\prime|)}{\xi} \right) 
\\
\nonumber
& 
+ \li \left( 1-\frac{\beta(E+|\vec{l}|)}{\xi} \right) 
- \li \left( 1-\frac{\beta(E'+|\vec{l}^\prime|)}{\xi} \right) 
\Bigg] \, ,
\end{align}
where $ll^\prime = m_\ell^2 + Q^2_\ell/2$ is the product of 
the 4-momenta of the incident and scattered lepton. The 
following abbreviations have been used:
\begin{align*}
& 
\beta = 
\frac{ll^\prime + \sqrt{(ll^\prime)^2 - m_\ell^4}}{m_\ell^2} 
\, , 
\\
& 
\xi = \frac{\beta ll^\prime - m_\ell^2}{\beta E - E'} \, .
\end{align*}
The non-radiative relative correction at first order for the 
cross section with no observed photon is IR finite and is given by
\begin{equation} 
\label{eq_delta_1}
\delta^{(1)}_{\text{non-rad}}(\Delta) 
= 
\delta^{(1)}_{1-\text{loop}} 
+ \delta^{(1)}_{1s\gamma}(\Delta) \, .
\end{equation}

\subsection{One hard photon cross section}

The cross-section for the radiative process with one hard 
photon is given by
\begin{equation}
\mathrm{d}^4 \sigma^{(1)}_{1h\gamma} 
= 
\frac{\mathrm{d}^4\Gamma _{1\gamma}}
{4 M |\vec{l}\, |} \overline{|\mathcal{M}_{1\gamma}|^2} 
\end{equation}
where the flux factor is given for the fixed-target frame 
and the bar indicates that one has to average and sum over 
the polarization degrees of freedom in the initial and final 
state, respectively. The differential phase-space is given by
\begin{align}
\nonumber
\mathrm{d}^4\Gamma_{1\gamma} 
=& 
\int \frac{1}{(2 \pi)^5} \, 
\mathrm{d}^4 l^\prime \, 
\mathrm{d}^4 k \, 
\mathrm{d}^4 p^\prime \, 
\delta (l^{\prime \, 2} - m_\ell^2) 
\delta (k^2) \delta (p^{\prime \, 2} - M^2) 
\delta^4 \left( l + p - l^\prime - p^\prime - k \right) 
\, .
\end{align}
We can choose a phase space parametrization in terms of 
energies and polar angles of the lepton and photon in the 
final state\footnote{There are alternative choices, for 
  example replacing the photon energy in favor of its 
  azimuthal angle, see e.g.~Ref.~\cite{Gramolin:2014pva}. 
  We have implemented also this option in our program for 
  numerical evaluations. We found excellent agreement between 
  the different phase space parametrizations, but one or the 
  other may be preferable for the implementation of kinematic 
  cuts depending on the experimental situation.} 
as described in detail in App.~\ref{app_1g-phase-space}. 
Using the notation defined there, the cross-section for one 
hard radiated photon becomes 
\begin{align}
\nonumber
\sigma^{(1)}_{1h\gamma} 
&= 
\frac{1}{32 (2 \pi)^4 M 
|\vec{l}\, |} \int \limits_{E'_\text{min}}^{E'_\text{max}}  
\mathrm{d} E' 
\int \limits_{ \theta_\ell ^\text{min}}^{ \theta_\ell^\text{max}} 
\mathrm{d} \cos \theta_\ell 
\int \limits_{\Delta}^{E_\gamma^\text{max}} 
\mathrm{d} E_\gamma 
\int \limits_{\theta_\gamma^\text{min}}^{ \theta_\gamma^\text{max}} 
\mathrm{d} \cos \theta_\gamma 
\\
\label{sigma_brems}
& 
\times \frac{ \overline{|\mathcal{M}_{1\gamma}|^2} }{\sin \theta_\ell 
\sin \theta_\gamma \sin \phi_\gamma} 
\Theta \left(1 - \frac{A^2}{B^2} \right) \, ,
\end{align}
where $\sin \phi_\gamma = \sqrt{1 - A^2/B^2}$. Explicit 
expressions for $A = A(E', \theta_\ell, E_\gamma, \theta_\gamma)$, 
$B = B(E', \theta_\ell, E_\gamma, \theta_\gamma)$ and the 
integration limits are given in App.~\ref{app_1g-phase-space}, 
see Eqs.~(\ref{Eq:AandB}) and (\ref{Eq:1hg-limits}). 
The matrix element squared is calculated with the help of 
the Feyncalc package and the final result is expressed in 
terms of invariant products of 4-momenta. A compact 
expression is given in App.~\ref{app_me}. We perform the 
numerical integration with the Cuba package \cite{Hahn:2004fe}.

\subsection{Vacuum polarization}
\label{sec_vac_pol}

The vacuum polarization, re-summed to all orders, leads to 
the replacement of the photon propagator, in Feynman gauge, 
by 
\begin{equation}
G_{\mu\nu} 
= \frac{-ig_{\mu \nu}}{q^2 \left[1 - \Pi (q^2) \right]} 
\, .
\end{equation}
The correction can be absorbed in the fine-structure constant 
\begin{equation} 
\label{eq_alpha_eff}
\alpha_\text{eff} (q^2) 
= \frac{\alpha}{1 - \Pi(q^2)}.
\end{equation}
The contribution from lepton loops is given at first order by
\begin{equation}
\delta_\text{vac-pol}^{\text{leptons}} 
= 2 \Pi_{e+\mu+\tau}(q^2) 
= \delta_\text{vac-pol}^e + \delta_\text{vac-pol}^\mu 
+ \delta_\text{vac-pol}^\tau 
\end{equation}
and can be written, for space-like momentum transfer $- q^2 
= Q^2 > 0$, in the compact form 
\begin{equation}
\delta_\text{vac-pol}^{\ell} 
= \frac{2\alpha}{3\pi} 
\left[ \left(v^2 - \frac{8}{3} \right) 
+ v \frac{(3-v^2)}{2} \ln \left(\frac{v+1}{v-1} \right) \right], 
\end{equation}
where $v = \sqrt{1 + 4 m_\ell^2/Q^2}$ with $m_\ell$ the mass 
of the lepton in the loop. At large $Q^2$ and including the 
two-loop contribution, one may use \cite{Kallen:1955fb}: 
\begin{equation}
\delta_\text{vac-pol}^{\ell} 
= 2 \frac{\alpha}{\pi} 
\left( \frac{1}{3} \ln \frac{Q^2}{m_\ell^2} - \frac{5}{9} \right) 
+ 2 \left( \frac{\alpha}{\pi}\right)^2 
\left( \frac{1}{4} \ln \frac{Q^2}{m_\ell^2} 
+ \zeta(3) - \frac{5}{24} \right)  
+ O(\alpha^3) \, .
\end{equation}
\begin{figure}[b!]
\centering
\includegraphics[width=0.75\textwidth]{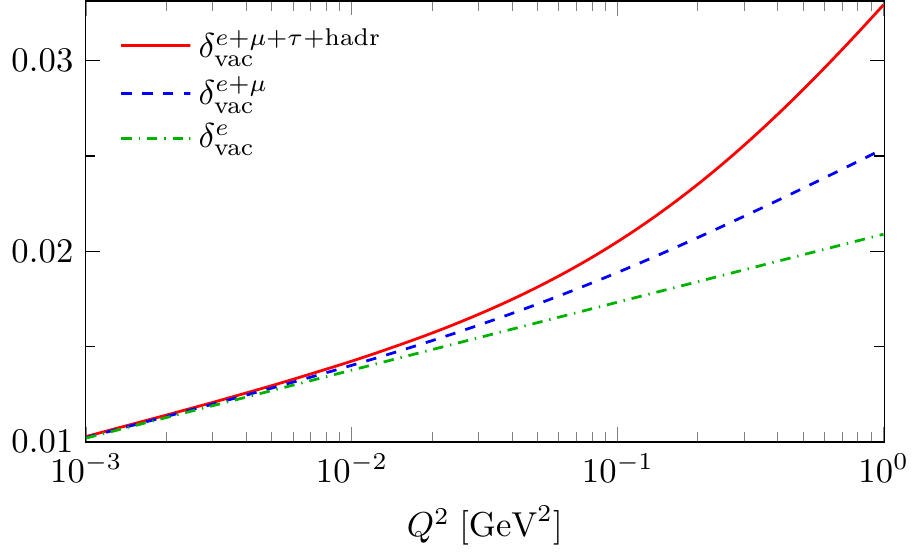}	
\caption{
Leptonic and hadronic contributions to the vacuum polarization 
in the $Q^2$ range relevant for the P2 experiment. 
}
\label{fig_delta_hadr}
\end{figure} 

The hadronic part of $\Pi(q^2)$ can be extracted from 
experimental data for the cross section of $e^+ e^-$ 
annihilation into hadrons. We use a table provided by 
F.~Ignatov~\cite{ignatov-vpl} (see also \cite{Actis:2010gg})
but it is straightforward to replace this by other 
parametrisations, as for example the one of 
Ref.~\cite{Jegerlehner:2011mw} or \cite{Keshavarzi:2018mgv}. 
In Fig.~\ref{fig_delta_hadr} 
we show numerical results for $\delta _\text{vac-pol}$. 
We conclude that one has to include the vacuum polarisation 
effect in a high-precision calculation of the cross section 
and contributions from other than electron loops should not 
be neglected for $Q^2$ values above a few times 
$10^{-2}$~GeV$^{2}$.

\section{Second-order corrections}
\label{sec_second_order}

\subsection{Non-radiative corrections}
\label{sec_2sgamma_cs}

\begin{figure}[b]\centering
\includegraphics[width=0.8\textwidth]{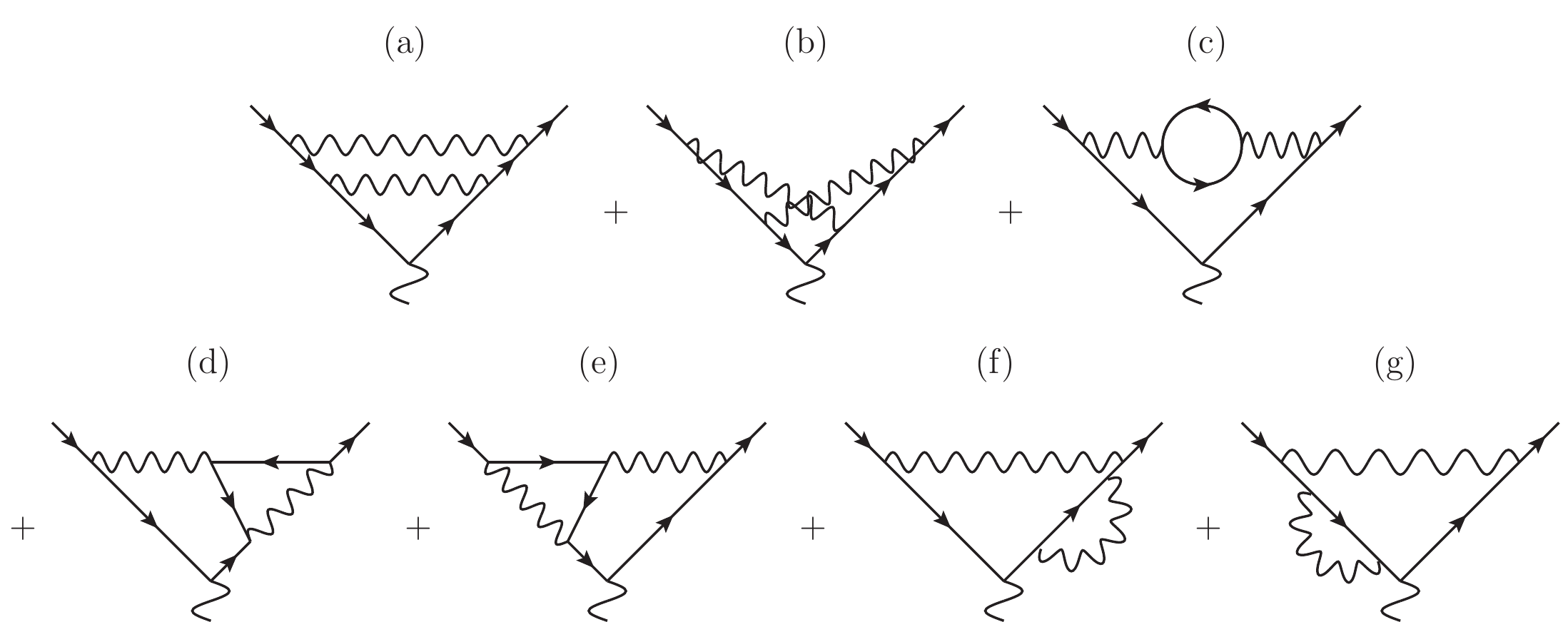}
\caption{
Feynman diagrams for two-loop vertex corrections. 
}
\label{diag_2-loop}
\end{figure}

The Feynman diagrams for two-loop corrections at the lepton 
line are shown in Fig.~\ref{diag_2-loop}. Their contribution 
to the matrix element is denoted by $\mathcal{M}_\text{2-loop}$. 
The relative two-loop correction factor includes the square 
of the one-loop corrections and is given by
\begin{equation}\label{eq_delta_2-loop}
\delta^{(2)}_{2-\text{loop}} = 
\frac{\left|\mathcal{M}_\text{vert}\right|^2
+ 2\Re(\mathcal{M}_\text{Born}^\dagger 
\mathcal{M}_\text{2-loop})}%
{\left|\mathcal{M}_\text{Born}\right|^2} 
\, .
\end{equation}
For electron scattering, the Pauli form factor $F_2^\ell$ 
can be neglected at this order. Then the two-loop correction 
reduces to
\begin{equation}
\delta^{(2)}_{2-\text{loop}} 
= 
\left(F_1^{\ell(1)}\right)^2 + 2 F_1^{\ell(2)}.
\end{equation}
A compact expression for $\delta^{(2)}_{2-\text{loop}}$, 
valid for $Q_\ell^2 \gg m_\ell^2$, can be extracted from  
Ref.~\cite{Hill:2016gdf} and is given by\footnote{We note 
  that the diagram of Fig.~\ref{diag_2-loop}c is taken 
  into account with an electron loop only. In principle, 
  there are also contributions with a heavy lepton or with 
  hadronic states in the loop. These contributions can be 
  calculated for example with the help of a dispersion 
  relation technique. From similar calculations for other 
  processes \cite{vanRitbergen:1998hn,Davydychev:2000ee}, 
  their numerical contribution can be estimated to be small. }
\begin{align}
\nonumber
\delta^{(2)}_{2-\text{loop}} 
=& 
\frac{1}{2}
\left(\delta_\text{1-\text{loop}}^{(1)}\right)^2 
+ \left(\frac{\alpha}{4\pi}\right)^2 
\Bigg[-\frac{8}{9}L^3 + \frac{76}{9}L^2 
+ \left(-\frac{979}{27}-\frac{44\pi^2}{9}+48\zeta(3)\right)L 
+\frac{4252}{27}
\\
& 
+ \frac{47\pi^2}{3}-16\pi^2\ln(2)-72\zeta(3)-\frac{64\pi^4}{45}
+ \mathcal{O}\left(\frac{m_\ell^2}{Q_\ell^2}\right) \Bigg] \, , 
\label{Eq:delta-2loop}
\end{align}
where $L=\ln(Q_\ell^2/m_\ell^2)$. Eq.~(\ref{Eq:delta-2loop}) 
agrees with the earlier calculation of 
Ref.~\cite{Mastrolia:2003yz}\footnote{The two-loop electron 
form factor from a calculation where both UV and IR divergences 
are isolated in dimensional regularization can be found 
in Ref.~\cite{Bonciani:2003ai}}. After removing the UV 
divergent parts the expression still contains IR divergences 
which cancel when soft-photon corrections are included at the 
level of the cross section. The soft-photon corrections at 
second order are corrections from two-soft-photon radiation 
and one-loop corrections for one-soft-photon radiation. 

\begin{figure}[t]\centering
\includegraphics[width=0.7\textwidth]{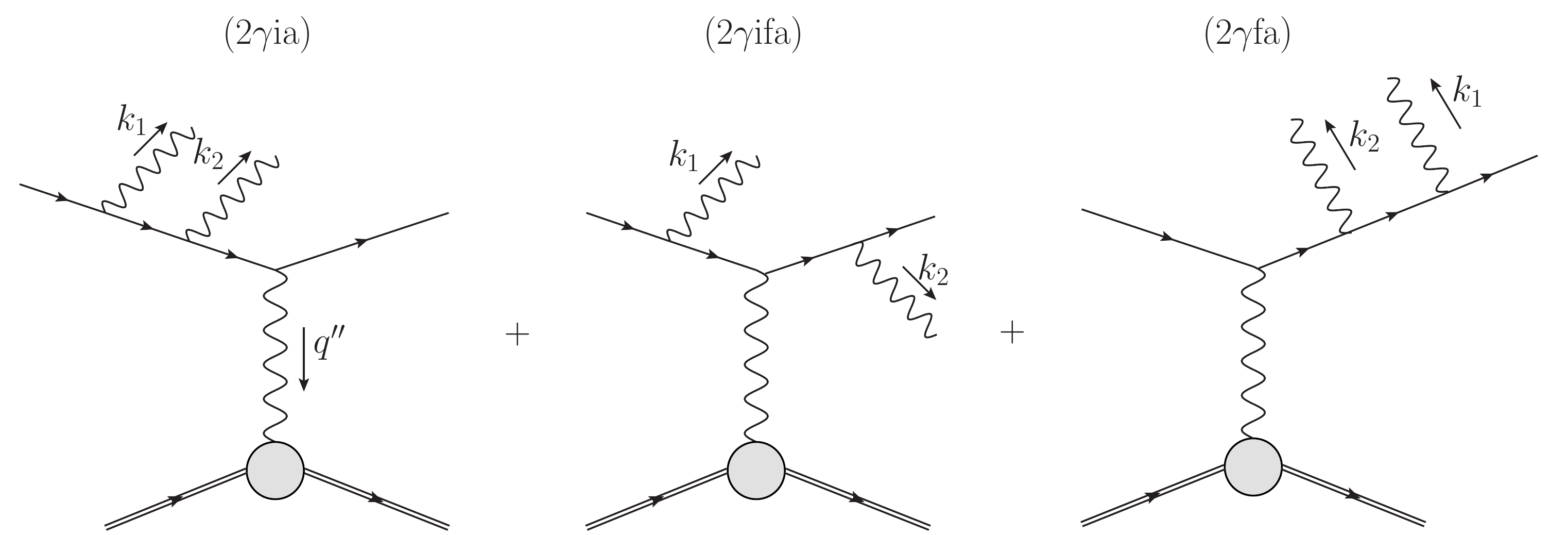}
\\[8mm]

\includegraphics[width=0.7\textwidth]{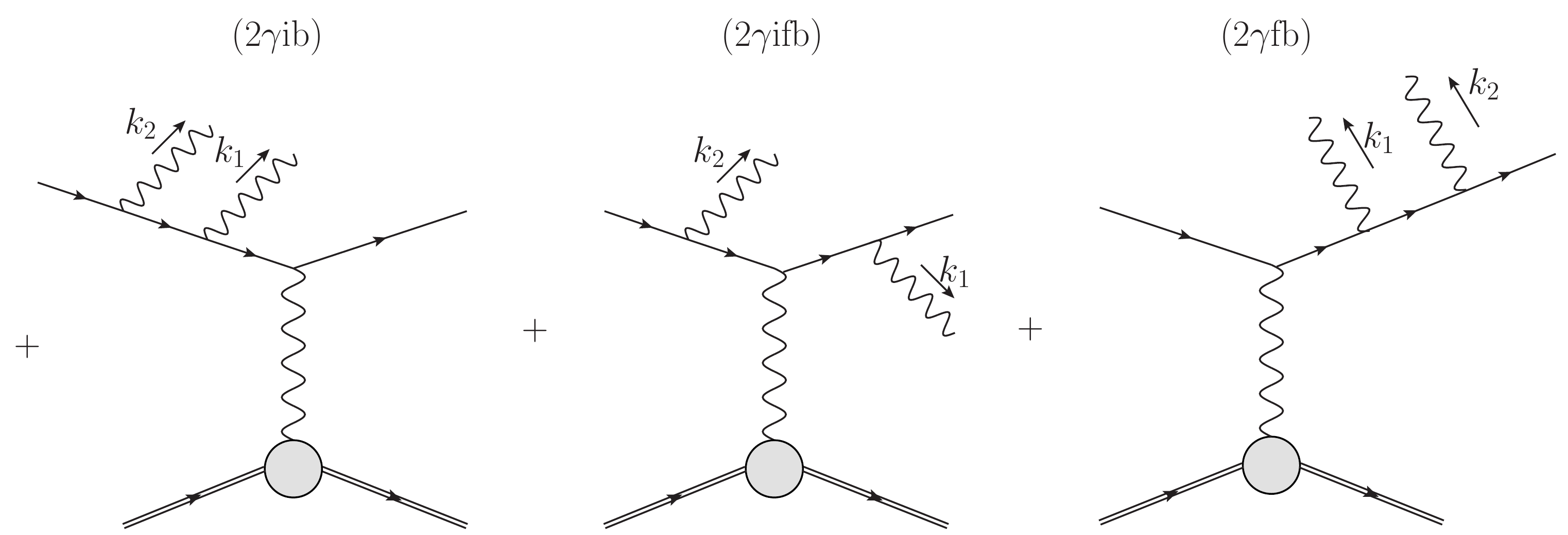}
\caption{Feynman diagrams for two-photon bremsstrahlung.}
\label{diag_2g}
\end{figure}

The diagrams for two-photon radiation are shown in 
Fig.~\ref{diag_2g}. In the soft-photon approximation, the 
corresponding correction factor is given by
\begin{equation} 
\label{eq_delta_2g_soft}
\delta^{(2)}_{2s\gamma} 
= 
\frac{1}{2!}\left(-\frac{\alpha}{4\pi^2}\right)^2
\int_{E_\gamma,\; E'_\gamma < \Delta} 
\frac{\mathrm{d}^3k_1}{E_\gamma}
\frac{\mathrm{d}^3k_2}{E'_\gamma}
\left(\frac{l_1}{l_1k_1} - \frac{l_2}{l_2 k_1}\right)^2
\left(\frac{l_1}{l_1k_2} - \frac{l_2}{l_2 k_2}\right)^2 \, .
\end{equation}
If both photon energies separately are taken smaller than the 
cut-off value $\Delta$, as we assume here, the phase space  
integration factorizes and leads to
\begin{equation}
\delta^{(2)}_{2s\gamma} 
= 
\frac{1}{2!} \left(\delta^{(1)}_{1s\gamma}\right)^2 \, .
\end{equation}
In contrast, if the integration is done by restricting the 
total unobserved energy, i.e., using $E_\gamma + E'_\gamma 
< \Delta$, as was done for example in \cite{Hill:2016gdf}, the 
soft-photon correction factor $\delta^{(2)}_{2s\gamma}$ 
contains an additional term $-\frac{\alpha^2}{3}(L-1)^2$, 
which comes from the phase-space overlap of the two photons 
\cite{Arbuzov:2015vba}. In our approach we take account of 
this overlap region in the contribution from two hard photons. 
Of course, the final result has to be independent of the way 
the phase-space slicing is implemented.

The Feynman diagrams for one radiated photon at one-loop 
order are shown in Fig.~\ref{diag_1g_1loop}. When treating 
the photon as soft, one can approximate their contribution 
by a factorized form in terms of the one-photon and one-loop 
correction factors: 
\begin{equation} 
\delta^{(2)}_{1-\text{loop}+1s\gamma}(\Delta) 
= 
\delta_{1-\text{loop}}^{(1)}\delta_{1s\gamma}^{(1)}(\Delta) 
\, .
\end{equation}
Combining all non-radiative second order corrections at the 
level of the cross section we obtain an IR finite, but cut-off 
dependent result, 
\begin{equation}
\delta^{(2)}_{\text{non-rad}}(\Delta) 
= 
\delta^{(2)}_{2-\text{loop}} 
+ \delta^{(1)}_{1-\text{loop}} \delta^{(1)}_{1s\gamma}(\Delta) 
+ \delta^{(2)}_{2s\gamma}(\Delta) 
\, .
\end{equation}

\subsection{One-loop corrections to radiative scattering}
\label{sec_1sgamma_1hgamma_cs}
 
\begin{figure}[b!]
\centering
\includegraphics[width=0.8\textwidth]{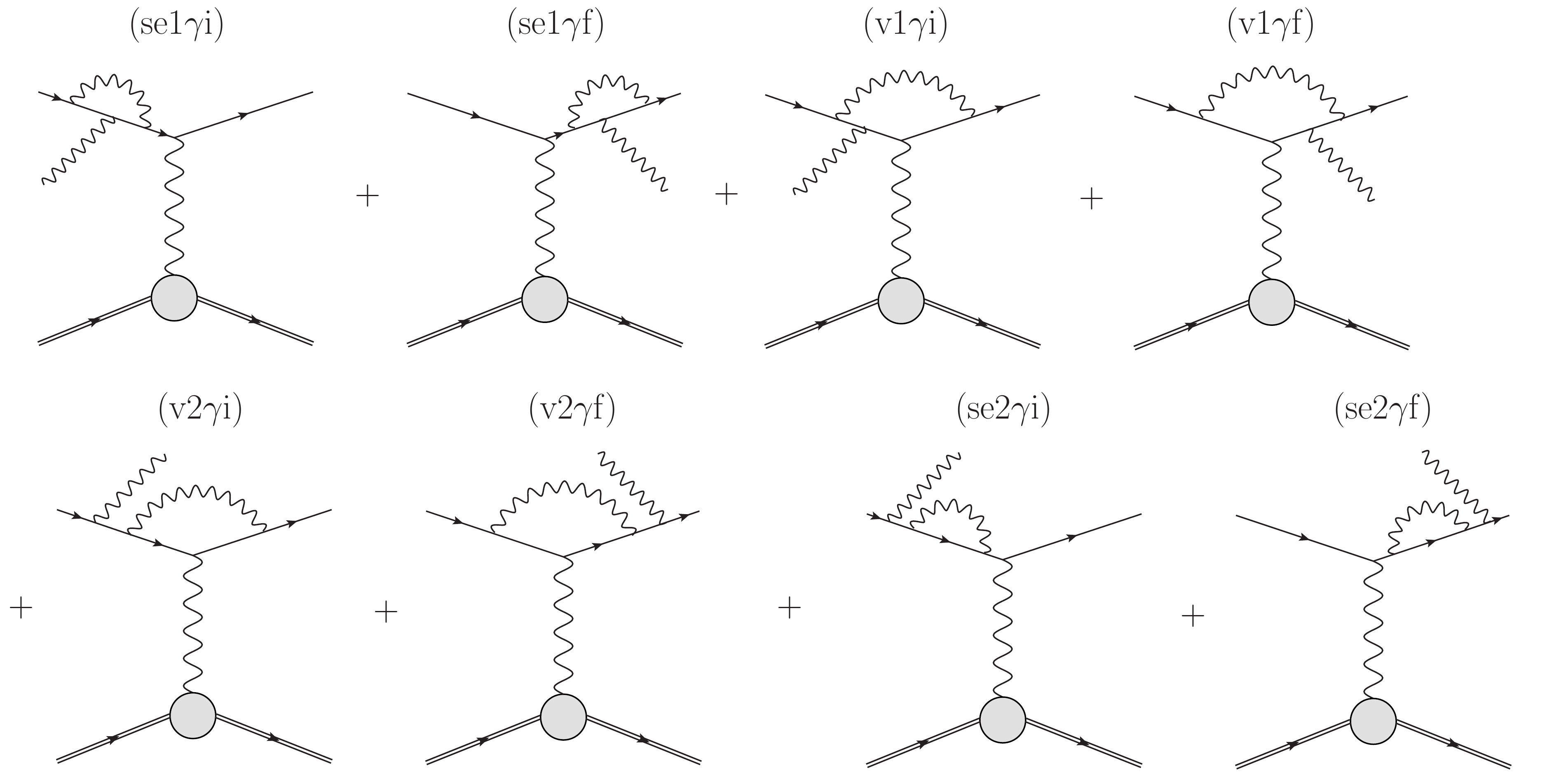}
\caption{Feynman diagrams for one-loop corrections to 
one-photon bremsstrahlung.}
\label{diag_1g_1loop}
\end{figure}

A complete second-order calculation of the cross section 
includes one-loop corrections for the process with one 
radiated photon. The corresponding Feynman diagrams are shown 
in Fig.~\ref{diag_1g_1loop}. Lepton self energy corrections 
at external legs are not needed since their contribution 
vanishes after renormalization as mentioned above. 

First, there are two self-energy diagrams with a photon 
radiated from the off-shell line. Their contribution to the 
matrix element is denoted by
\begin{equation}
\mathcal{M}_{\text{se1}\gamma} 
= 
\mathcal{M}_{\text{se1}\gamma \text{i}} 
+ 
\mathcal{M}_{\text{se1}\gamma \text{f}} \, .
\end{equation}
The one-loop integral entering $\mathcal{M}_{\text{se1}\gamma}$ 
is IR finite. Its UV divergence is removed by adding the vertex 
counter-term proportional to $\delta Z_1$ (see 
Eq.~(\ref{Eq:deltaZ1})). 

Next, two diagrams are vertex corrections with a photon emitted
from the off-shell line and their matrix element is denoted by
\begin{equation}
\mathcal{M}_{\text{v1}\gamma} 
= 
\mathcal{M}_{\text{v1}\gamma \text{i}} 
+ 
\mathcal{M}_{\text{v1}\gamma \text{f}} \, .
\end{equation}
The 4-point one-loop integrals needed here are UV finite, 
but contain IR-divergent contributions.  

The second row of diagrams in Fig.~\ref{diag_1g_1loop} 
have a photon attached to an external, on-shell lepton 
line; two of them are self energy insertions in the 
off-shell lepton line,  
\begin{equation}
\mathcal{M}_{\text{se2}\gamma} 
= 
\mathcal{M}_{\text{se2}\gamma \text{i}} 
+ 
\mathcal{M}_{\text{se2}\gamma \text{f}} \, ,
\end{equation}
and two of them describe a one-loop vertex correction, 
\begin{equation}
\mathcal{M}_{\text{v2}\gamma} 
= 
\mathcal{M}_{\text{v2}\gamma \text{i}} 
+ 
\mathcal{M}_{\text{v2}\gamma \text{f}} \, .
\end{equation}
These diagrams are UV divergent and require renormalization 
by including counter-terms, either at the vertex 
($\delta Z_1$) or at the lepton self energy ($\delta Z_2$ 
and $\delta m_\ell$). 

The second-order corrections due to these diagrams are 
obtained from the interference with the first-order 
diagrams. They can be split into parts, 
\begin{equation}
\label{eq_delta_1-loop_2}
\delta^{(2)}_{1-\text{loop} + 1h\gamma} 
= 
\delta_{\text{se1}\gamma} 
+ \delta_{\text{v1}\gamma} 
+ \delta_{\text{se2}\gamma} 
+ \delta_{\text{v2}\gamma} 
\, ,
\end{equation}
with an obvious meaning of the indices as explained above.

\subsection{One hard and one soft photon}
\label{sec_1hg_1sg}

IR divergences contained in one-loop corrections to the 
radiative process described in the previous sub-section 
are cancelled by corresponding IR divergences from 
soft-photon contributions of 2-photon bremsstrahlung. 
The diagrams for $\ell p \rightarrow \ell p \gamma\gamma$
have been presented above in Fig.~\ref{diag_2g}. We 
separate IR divergent contributions by assuming one 
photon is soft and the other is hard. In the soft-photon 
approximation, the IR divergence can be factorized as 
described above. If the soft photon has momentum $k_1$, 
the IR part of $\mathcal{M}_{2\gamma}$ is contained in 
\begin{equation}
\mathcal{M}_{2\gamma}^\text{IR-div} = 
- e \mathcal{M}_{1\gamma}(k_2) 
\left(\frac{l \epsilon^*}{l k_1} -  
\frac{l^\prime \epsilon^*}{l^\prime k_1}\right) 
\, , 
\end{equation}
where $\mathcal{M}_{1\gamma}(k_2)$ is the matrix element 
for bremsstrahlung of a photon with momentum $k_2$. Of 
course, one has to include a similar term with interchanged 
photon momenta $k_1\leftrightarrow k_2$. 

The cross section for two photons in the final state is given 
by 
\begin{equation}
\label{eq_sigma_2g}
\mathrm{d}^7 \sigma_{2 \gamma} 
= 
\frac{\mathrm{d}^7\Gamma_{2\gamma}}{2 \cdot 4 M |\vec{l}|} 
\overline{|\mathcal{M}_{2\gamma}|^2} \, ,
\end{equation}
where a symmetrization factor of one-half is applied to take 
into account the fact that there are two identical particles in 
the final state. Making the integration over the photon energies 
explicit and taking into account that the cross section is 
symmetric with respect to interchanging $k_1 \leftrightarrow 
k_2$, the separation between soft- and hard-photon phase 
space regions is done as
\begin{equation}
\label{eq_sg_hg_separation}
\int_{0}^{E_\gamma ^\text{max}} \mathrm{d} E_{\gamma }
\int_{0}^{E_\gamma ^\text{max}} \mathrm{d} E_{\gamma}'  
= 
\int_{0}^{\Delta} \mathrm{d} E_{\gamma}
\int_{0}^{\Delta} \mathrm{d} E_{\gamma}'  
+ 2\int_{\Delta}^{E_\gamma ^\text{max}} \mathrm{d} E_{\gamma}
\int_{0}^{\Delta} \mathrm{d} E_{\gamma}'  
+ \int_{\Delta}^{E_\gamma ^\text{max}} \mathrm{d} E_{\gamma}
\int_{\Delta}^{E_\gamma ^\text{max}} \mathrm{d} E_{\gamma}' 
\, . 
\end{equation}
Only the second term on the right-hand side of  
Eq.~(\ref{eq_sg_hg_separation}) contributes to the part of 
the total cross section considered here, where we require 
one hard and one soft photon. The first part is purely 
soft-photon and contributes to the non-radiative cross section, 
combined with the 2-loop contribution. The last term for 
two hard photons is described in the next sub-section. 
 
The infrared divergence can be factorized, resulting in  
\begin{equation}
\label{eq_sg_hg_fact}
\sigma^{(2)}_{1s\gamma+1h\gamma}(\Delta)
= 
2 \int_{\Delta}^{E_\gamma^{\text max}} 
\int_0^\Delta 
\left(\mathrm{d}^7 \sigma_{2 \gamma}\right)_{1\gamma \rightarrow 0} 
=
\delta^{(1)}_{1s\gamma}(\Delta) \int_{\Delta}^{E_\gamma^{\text max}} 
\mathrm{d}^4 \sigma _{1 \gamma}
+ 2 \int_{\Delta}^{E_\gamma^{\text max}} 
\int_0^\Delta \left(\mathrm{d}^7 
\sigma_{2 \gamma}\right)_{1\gamma \rightarrow 0}^\text{IR-finite},
\end{equation} 
where $\delta^{(1)}_{1s\gamma}(\Delta)$ was defined in 
Eq.~(\ref{eq_delta_1sg}) and the factor of 2 was cancelled 
by the symmetry factor of the 2-photon cross section. Its 
IR-divergent part cancels against corresponding parts of 
the one-loop corrections to radiative scattering. One can 
show that the factorized part (the first term of the 
right-hand side of Eq.~(\ref{eq_sg_hg_fact})) emerges from 
those four Feynman diagrams of Fig.~\ref{diag_2g} where 
the soft photon is emitted from an on-shell line. The two 
remaining diagrams with a soft photon coming from an 
off-shell line lead to an IR-finite contribution (the 
last term in Eq.~(\ref{eq_sg_hg_fact})). This part can be 
safely calculated by numerical methods.

\subsection{Two hard-photons}
\label{sec_2hgamma_cs}

The cross section for $\ell p \rightarrow \ell p\gamma\gamma$ 
is given by Eq.~(\ref{eq_sigma_2g}). The differential phase-space 
is given by 
\begin{align}
\nonumber
\mathrm{d}^7\Gamma 
=& 
\int \frac{1}{(2 \pi)^8} \, 
\mathrm{d}^4 l^\prime \, 
\mathrm{d}^4 p^\prime \, 
\mathrm{d}^4 k_1 \, 
\mathrm{d}^4 k_2 \, 
\delta (l^{\prime \, 2} - m_\ell^2) 
\delta (p^{\prime \, 2} - M^2) 
\delta (k_1^2) \delta (k_2^2) 
\\ & 
\times 
\delta^4 \left( l + p - l^\prime - p^\prime - k_1 - k_2 \right) 
\, . 
\end{align}                   
The treatment of the delta-functions and the derivation of 
integration limits is described in App.~\ref{app_2g-phase-space}. 
The spin-averaged square of the matrix element is calculated 
with the Feyncalc package and expressed in terms of invariants.
The result is lengthy and not given here. Using the notation 
defined in App.~\ref{app_2g-phase-space}, the cross section for 
two-hard-photon radiation is expressed as
\begin{align}
\sigma^{(2)}_{2h\gamma}
&= 
\frac{E_\gamma E'_\gamma |\vec{l}^{\, \prime}|}%
{128 (2 \pi)^7 M |\vec{l}|} 
\int \limits_{E'_\text{min}}^{E'_\text{max}} \mathrm{d} E' 
\int \limits_{\cos\theta_{\ell,\text{min}}}^{\cos\theta_{\ell,\text{max}}} 
\mathrm{d} \cos\theta_\ell 
\int \limits_{\Delta}^{E_{\gamma,\text{max}}} \mathrm{d} E_\gamma 
\int \limits_{\Delta}^{E^\prime_{\gamma,\text{max}}} 
\mathrm{d} E'_\gamma 
\\
\nonumber
& \times 
\int 
\limits_{\cos \theta^\prime_{\gamma,\text{min}}}^{\cos \theta^\prime_{\gamma,\text{max}}}  
\mathrm{d} \cos \theta'_\gamma 
\int 
\limits _{\phi^\prime_{\gamma,\text{min}}}^{\phi^\prime_{\gamma,\text{max}}}  
\mathrm{d} \phi^\prime_\gamma 
\int 
\limits_{\cos \theta_{\gamma,\text{min}}}^{\cos \theta_{\gamma,\text{max}}}  
\mathrm{d} \cos \theta_\gamma 
\frac{\overline{|\mathcal{M}_{2\gamma}|^2}}%
{\left|\alpha_1 \cos \phi_\gamma - \alpha_2 \sin \phi_\gamma \right|} 
\Theta \left(1 - \frac{\alpha_3^2}{\alpha_1^2 + \alpha_2 ^2} 
\right) \, . 
\end{align}
The integration limits and the definition of the quantities 
$\alpha_i$ are given in App.~\ref{app_2g-phase-space}. 
Since the IR poles are cut off by lower integration limits 
on the photon energies, one can, in principle use standard 
integration packages for numerical calculations. It turns 
out, however, that collinear poles in the differential cross 
section render a naive approach numerically unstable. In 
order to deal with this problem we have used a partial 
fractioning to separate the collinear poles
\begin{equation} 
\label{eq_col_poles} 
\mathrm{d}^7 \sigma^{(2)}_{2 \gamma} 
\propto 
\frac{1}{(lk_1) \cdot (lk_2) \cdot (l^\prime k_1) 
\cdot (l^\prime k_2)} 
\rightarrow 
\frac{A}{lk_1} + \frac{B}{lk_2} 
+ \frac{C}{l^\prime k_1} + \frac{D}{l^\prime k_2} 
\, . 
\end{equation}
For each term in the sum after partial fractioning we 
find a specific change of integration variables which 
allows us to obtain an efficient and numerically stable 
integration.

\section{Numerical tests}
\label{sec_tests}

The presence of divergences in intermediate results forces 
us to introduce various regularization parameters which must 
cancel in the final result: 
\begin{itemize}
\item 
UV divergences are treated in dimensional regularization 
where pole terms in $\Delta_\epsilon$ appear. They 
are accompanied by logarithms of a mass scale parameter 
$\mu$ introduced to keep the mass dimension of loop 
integrals homogeneous. Both the $\Delta_\epsilon$- and 
$\mu$-dependence cancel by including corresponding 
counter-terms. 
\item 
IR divergences are regularized by a finite photon mass 
$\lambda$. Logarithms of the photon mass have to cancel 
exactly between loop contributions and corrections from 
soft-photon radiation. 
\item 
The phase space slicing parameter $\Delta$ was introduced 
to separate soft-photon from hard-photon contributions. 
The part with $E_\gamma, E^\prime_\gamma < \Delta$ is 
calculated in the soft-photon approximation. Therefore 
the $\Delta$-dependence disappears only in the limit 
$\Delta \rightarrow 0$. A residual $\Delta$-dependence 
may be visible if $\Delta$ is chosen too large. 
\end{itemize}

\begin{table}[t]
\centering
\begin{tabular}{l | l | l | l | l }
	\hline
	\rule{0pt}{5mm}
	$\lambda^2$ [GeV$^2$]& $m_e^2$ & $10^{-4}$ & $10^{-8}$ & 
	$10^{-12}$ 
	\\
	\hline
	\rule{0pt}{5mm}
	$\sigma^{(2)}_{1h\gamma}$ & $4034.10 \pm 0.92$ & 
	$4033.33 \pm 0.92$ & $4033.89 \pm 0.87$ & $4033.79 \pm 0.88$ 
	\\
	\hline
	\multicolumn{5}{c}{\vspace{0.2cm}} 
	\\
	\hline
	\rule{0pt}{5mm}
	$\Delta_\epsilon$ & $0$ & $10^2$ & $10^4$ & $10^{6}$ 
	\\
	\hline
	\rule{0pt}{5mm}
	$\sigma^{(2)}_{1h\gamma}$ & $4034.10 \pm 0.92$ &
	$4033.84 \pm 0.91$ & $4033.41 \pm 0.90$ & $4031.11 \pm 0.92$ 
	\\
	\hline
	\multicolumn{5}{c}{\vspace{0.2cm}} 
	\\
	\hline
	\rule{0pt}{5mm}
	$\mu ^2$ & $1$ & $10^4$ & $10^{8}$ & $10^{12}$ 
	\\
	\hline
	\rule{0pt}{5mm}
	$\sigma^{(2)}_{1h\gamma}$ & $4034.10 \pm 0.92$ &
	$4033.03 \pm 0.90$ & $4034.04 \pm 0.94$ & $4033.41 \pm 0.92$	
\end{tabular} 
\caption{
Test of the independence of $\sigma^{(2)}_{1h\gamma}$ 
(see Eq.~(\ref{eq:sigma2-1gh})) on the unphysical 
regularization parameters $\lambda^2$, $\Delta_\epsilon$ 
and $\mu ^2$. The uncertainty estimates are due to the 
finite statistics of the Monte Carlo integration. 
}
\label{table_test_delta_1loop_1g}
\end{table}

\begin{figure}[t]
\centering
\includegraphics[width=0.45\textwidth]{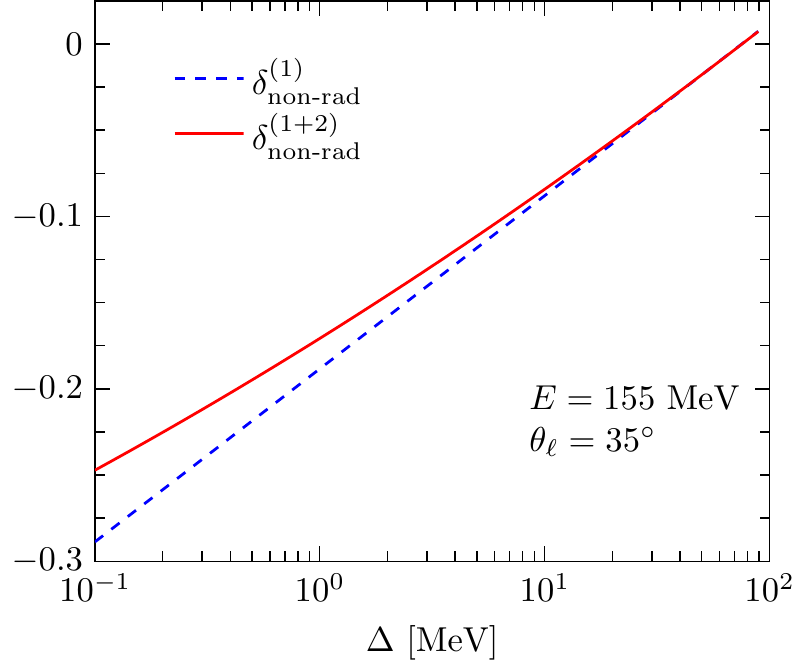}
\caption{
The dependence of the non-radiative parts of the correction 
factors at first and second order on the phase space slicing 
cut-off $\Delta$. 
}
\label{fig_delta_fwd}
\end{figure} 

The implementation of loop integrals in the LoopTools 
package \cite{Hahn:1998yk} allows us to keep the 
parameters $\Delta_\epsilon$, $\mu$ and $\lambda$ 
in separate parts of the calculation. Their cancellation 
can therefore be tested numerically. As an example, we 
show numerical results for the $O(\alpha)$ corrected 
one-photon bremsstrahlung cross section, given in 
Eq.~(\ref{eq:sigma2-1gh}) for P2 kinematics, i.e.\ 
for $E = 155$~MeV, $\theta_\ell = 35^\circ \pm 10^\circ$ and 
$E'_\text{min} = 45$~MeV and with a cut-off for the photon 
energy of $\Delta = 10$~MeV. We take default values for 
the three regularization parameters as $\lambda^2 = 
m_e^2$, $\Delta_\epsilon = 0$ and $\mu^2 = 1$. From 
Tab.~\ref{table_test_delta_1loop_1g} we conclude that 
these unphysical parameters can be varied over a very 
large range of values without leading to a significant 
numerical variation of the correction factor. The 
observed behaviour constitutes a test of an important 
part of the calculation. 

\begin{figure}[t!]
\centering
\includegraphics[width=0.8\textwidth]{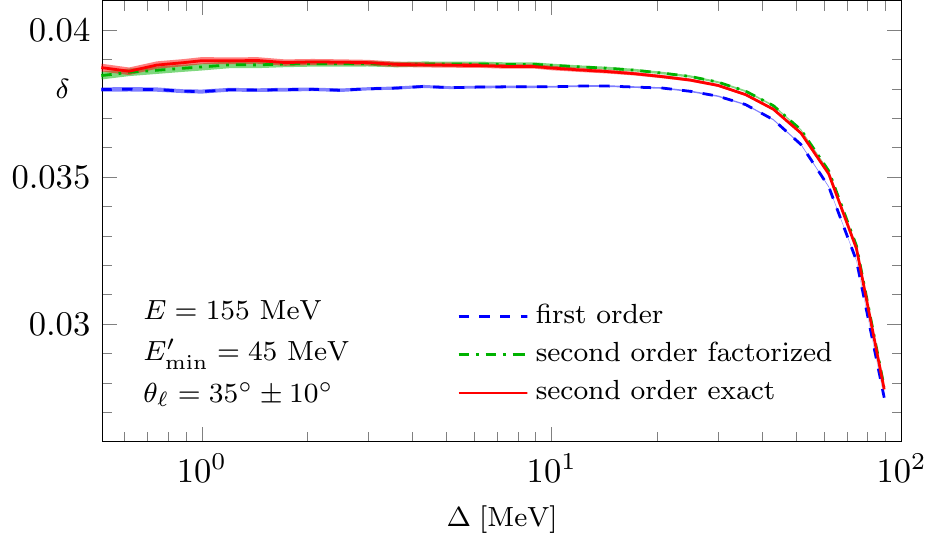}	
\caption{
Test of the $\Delta$-independence of the complete correction 
factors when non-radiative contributions and hard-photon 
radiative effects are added, at first order $\delta = 
\sigma^{(1)} / \sigma^{(0)} -1$ and at second order $\delta = 
\sigma^{(2)} / \sigma^{(0)} -1$. Beam energy and the range of 
the scattering angle is chosen for the P2 experiment. The 
meaning of the labels ``factorized'' and ``exact'' is 
explained in the text.
}
\label{fig_total_cs}
\end{figure} 

The non-radiative parts of the cross section depend 
logarithmically on the phase space slicing parameter $\Delta$, 
at first order $\delta^{(1)}_{\text{non-rad}} \propto 
\ln \Delta$, at second order $\delta^{(2)}_{\text{non-rad}} 
\propto \ln^2 \Delta$. A typical example of numerical results 
of the $\Delta$-dependence for the case of the P2 experiment 
is shown in Fig.~\ref{fig_delta_fwd}. This $\Delta$-dependence 
is cancelled by the hard-photon contribution. In 
Fig.~\ref{fig_total_cs} we show an example, again for the 
kinematics of the P2 experiment. At large values of the 
soft-photon cut-off, when $\Delta$ reaches some 10 percent 
of the beam energy, the break-down of the soft-photon 
approximation is visible. Below $\Delta \simeq 10$~MeV, 
there is a nice plateau where the total result is 
independent of the cut-off. At first order, the cancellation 
looks perfect while for the second-order calculation one 
can observe that the  numerical cancellation becomes less 
and less stable for decreasing $\Delta$. However, the choice 
$1~\text{MeV} \lsim \Delta \lsim 10$~MeV is appropriate for 
the P2 experiment and guarantees  that the soft-photon 
approximation used for the calculation of the non-radiative 
part of the total correction does not lead to a significant 
distortion of the total result. 

It is also interesting to study an approximation for the 
calculation of $\delta^{(2)}_{1-\text{loop} + 1h\gamma}$, 
Eq.~(\ref{eq_delta_1-loop_2}). The approximation consists in 
assuming that the one hard-photon correction and the loop 
correction factorize also for finite values of the photon 
energy. This amounts to the replacement 
\begin{equation}
\delta^{(2)}_{1-\text{loop} + 1h\gamma} 
\rightarrow 
\delta^{(1)}_{1-\text{loop}}.
\end{equation} 
Figure~\ref{fig_total_cs} shows an example of numerical 
results based on this approximation (green, dash-dotted 
line). We find good agreement between the exact calculation 
and the approximation, at the level of $10^{-4}$ and better 
for the energy and angle range shown in this figure. In 
a more detailed study we found that the differences are 
largest in the vicinity of the final-state radiation peak.

\section{Numerical results}
\label{sec_results}

We start this section with the discussion of a few numerical 
results for leptonic radiative corrections which are relevant 
for the P2 experiment at the MESA facility in 
Mainz~\cite{Becker:2018ggl}. The P2 experiment plans to 
measure the parity-violating asymmetry in elastic electron 
proton scattering with a polarized electron beam of energy 
$E = 155$~MeV. The P2 spectrometer covers an angular acceptance 
range of $35 \pm 10^\circ$ and for simplicity we assume that 
only scattered electrons with a fixed energy of at least 
$E^\prime_{\rm min} = 45$~MeV are detected. The average 
momentum transfer squared is $\langle Q^2 \rangle = 6 \cdot 
10^{-3}$~GeV$^2$. An ancillary measurement for the 
determination of the axial and strange magnetic form factors 
at backward angles is also possible. Such a measurement could 
cover the angular range $135^\circ \leq \theta_\ell \leq 
155^\circ$. We repeat that we have used a simple dipole 
parametrization for the proton form factors, $G_E = 
\left(1 + Q^2 / \Lambda^2\right)^{-2}$ and $G_M = 
\kappa_p G_E$ with $\Lambda = 0.71$~GeV$^2$ and we have 
checked that, while the cross sections can change by 
a few per mill when using a different form factor 
parametrization, the correction factors are insensitive 
to this choice at the level well below one per mill. 

\begin{figure}[t!]
\centering
\includegraphics[width=0.48\textwidth]{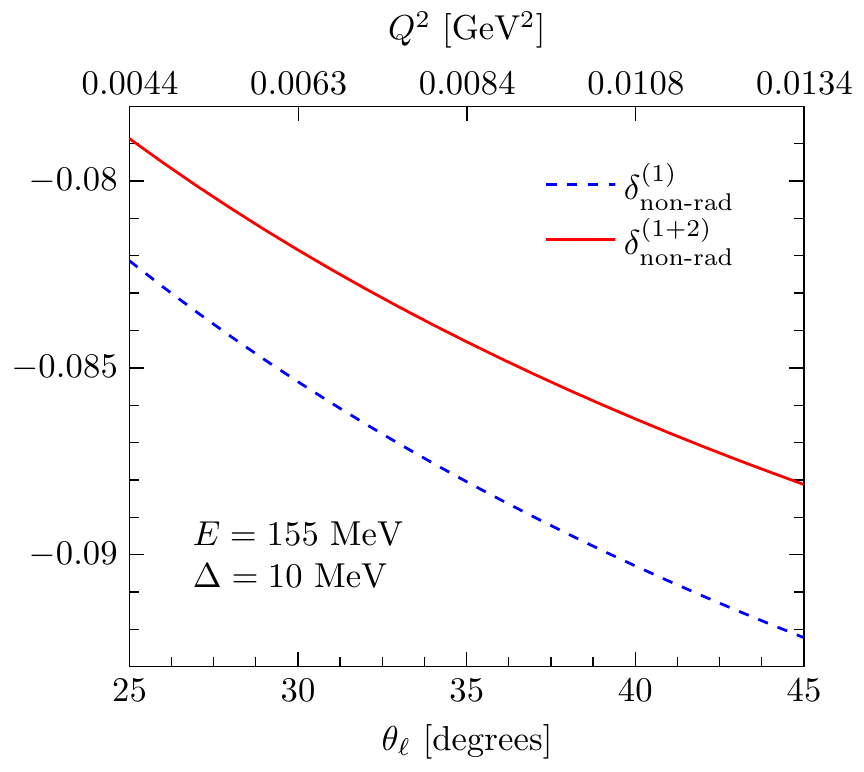}	
\includegraphics[width=0.48\textwidth]{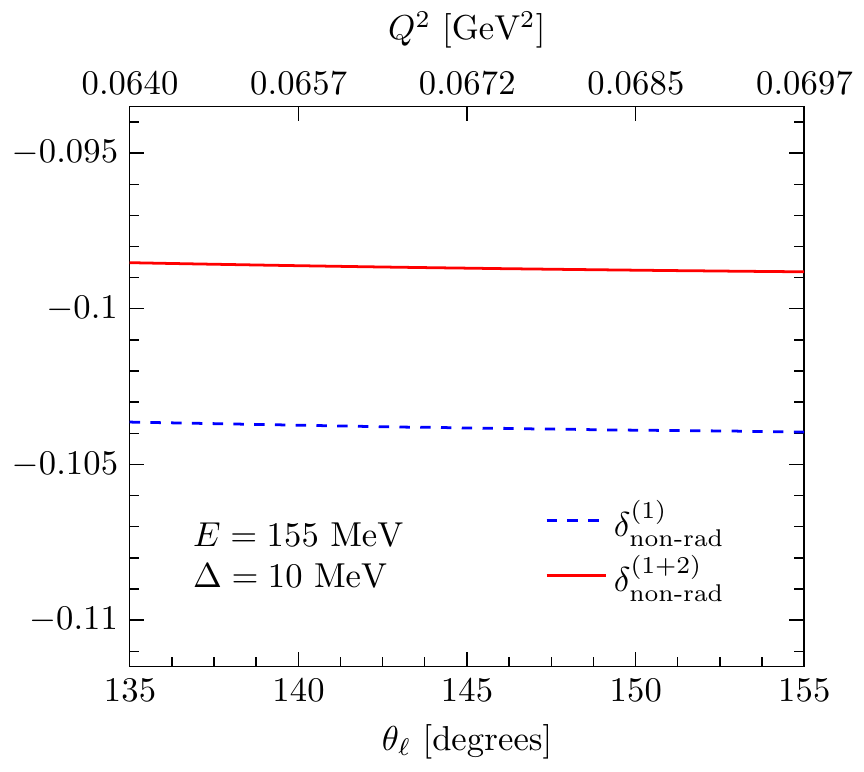}	
\caption{
The non-radiative part of the correction factors at first 
(dashed, blue curve) and second order (full, red curve) 
in the range of scattering angles relevant for P2: forward 
scattering with $25^\circ < \theta_\ell < 45^\circ$ (left) and 
backward scattering with $135^\circ < \theta_\ell < 155^\circ$ 
(right). Soft-photon corrections are included with the cut-off 
fixed at $\Delta = 10$~MeV.    
}
\label{fig_delta_nonrad}
\end{figure} 
\begin{figure}[b!]
\centering
\includegraphics[width=0.45\textwidth]{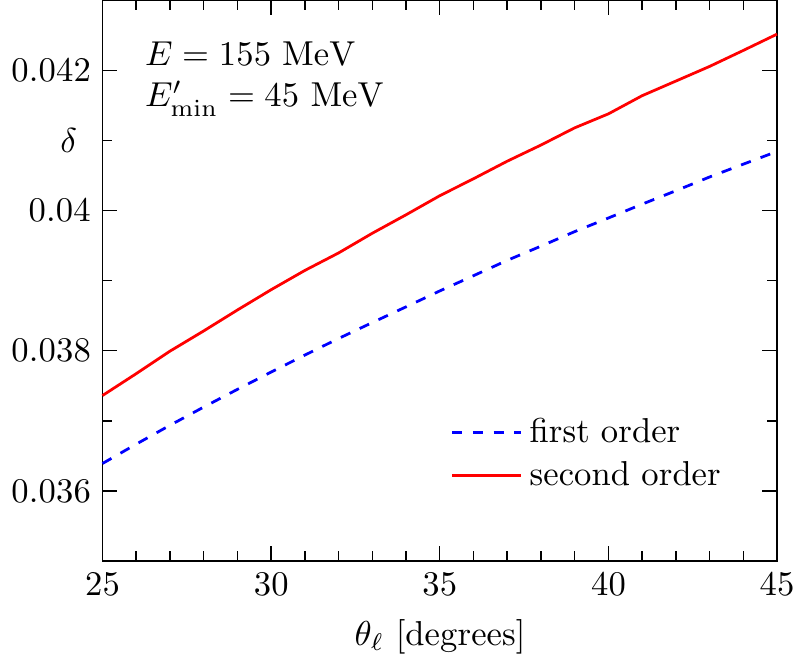}	
\includegraphics[width=0.45\textwidth]{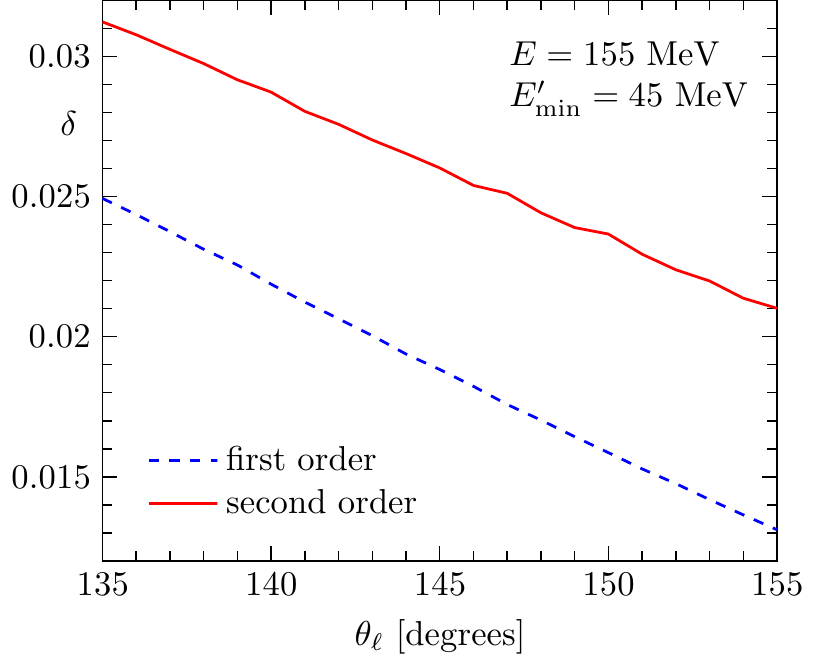}	
\caption{
The complete radiative correction factors at first 
(dashed, blue curve) and second order (full, red curve) 
in the range of scattering angles relevant for P2: forward 
scattering with $25^\circ < \theta_\ell < 45^\circ$ (left) and 
backward scattering with $135^\circ < \theta_\ell < 155^\circ$ 
(right). Hard-photon radiation is included with the 
restriction that the final state electron has an energy 
above $E^\prime_{\rm min} = 45$~MeV.    
}
\label{fig_delta_tot_P2}
\end{figure} 

The non-radiative part of the corrections is shown in 
Fig.~\ref{fig_delta_nonrad}. In this case, soft-photon 
radiation is included with the cut-off $\Delta = 10$~MeV. 
The corrections reach the level of some $-8.5\,\%$ and 
exhibit a moderate $Q^2$ dependence. Second-order 
corrections are small, but are relevant at the level 
of $0.3$ to $0.5\,\%$. The layout of the P2 spectrometer 
with a solenoidal magnetic field is constructed in such 
a way that no bremsstrahlung photon emitted in the 
target volume can reach the detector. Therefore, radiative 
scattering will contribute to the measured cross section 
as long as the scattered electrons fulfil the condition 
$E^\prime > E^\prime_{\rm min}$. The complete radiative 
correction factor including hard-photon radiation is 
shown in Fig.~\ref{fig_delta_tot_P2}. We find that the 
cross section is increased significantly by the inclusion 
of radiative processes. The corrections are now positive, 
at the level of a few percent. The difference between 
the first- and second-order calculations turns out to 
be slightly smaller in the forward region, but can still 
reach somewhat more than half a percent in the backward 
region. 

\begin{figure}[t!]
\centering
\includegraphics[width=0.45\textwidth]{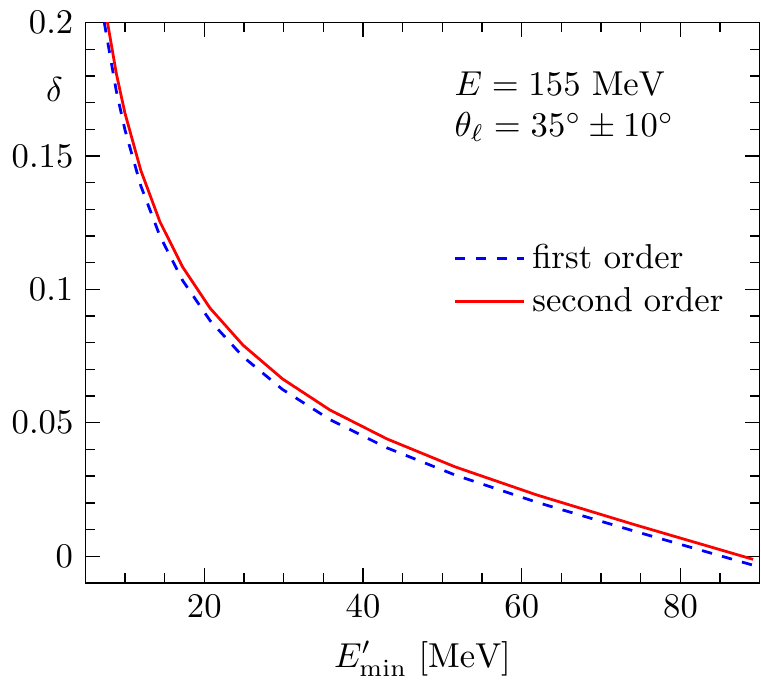}
\includegraphics[width=0.485\textwidth]{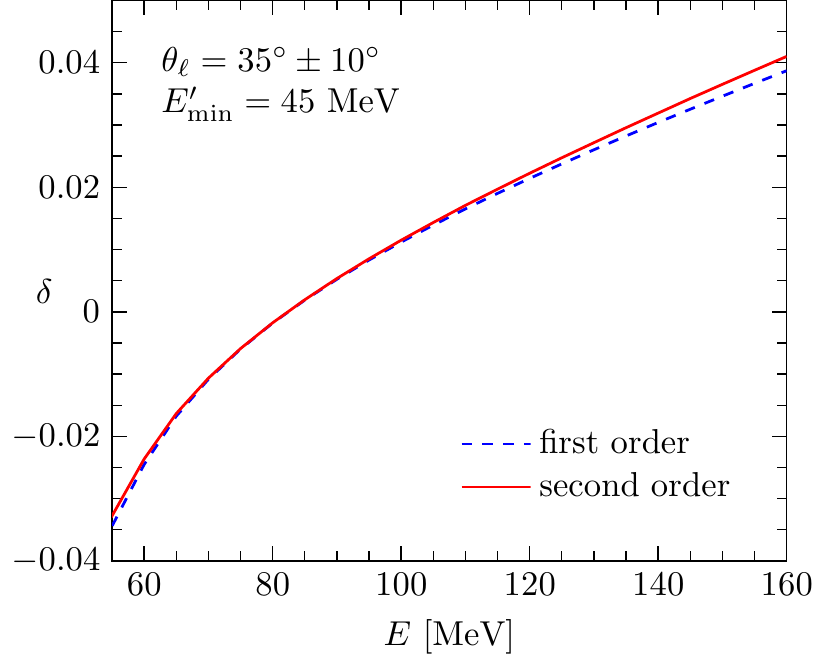}
\caption{
The complete radiative correction factors at first 
(dashed, blue curve) and second order (full, red curve) 
for P2 as a function of the cut-off on the energy of 
the scattered electron $E^\prime_{\rm min}$ (left) and 
as a function of the beam energy with fixed $E^\prime_{\rm min}$ 
(right). The scattering angle was integrated over the range 
$25^\circ \leq \theta_\ell \leq 45^\circ$. 
}
\label{fig_delta_tot_E1min}
\end{figure}

While it is not possible at the P2 experiment to impose 
a veto on hard radiated photons directly, the requirement 
of a minimum energy $E^\prime_{\rm min}$ for the scattered 
electrons restricts the phase space for photon emission 
indirectly. This introduces a strong dependence on 
$E^\prime_{\rm min}$. Numerical results are shown in 
Fig.~\ref{fig_delta_tot_E1min} (left), again both at first 
and at second order. In the P2 experiment energy loss can 
also occur when the incoming electron passes through the 
liquid hydrogen target. It is therefore also important to 
know how the cross section depends on the energy $E$ of the 
incoming electrons. Results for the first- and second-order 
radiative correction factors are shown in the right part of 
Fig.~\ref{fig_delta_tot_E1min}. For reduced $E$ while keeping 
$E^\prime_{\rm min} = 45$~MeV fixed, hard-photon radiation 
will be suppressed and the corrections become negative. The 
observed strong dependence on $E$ and $E^\prime_{\rm min}$ 
highlights the necessity to include radiative effects in a 
full Monte-Carlo simulation of the experiment where the 
acceptance for electron detection may be a complicated 
function of the scattering angle. 

\begin{figure}[t!]
\centering
\includegraphics[width=0.9\textwidth]{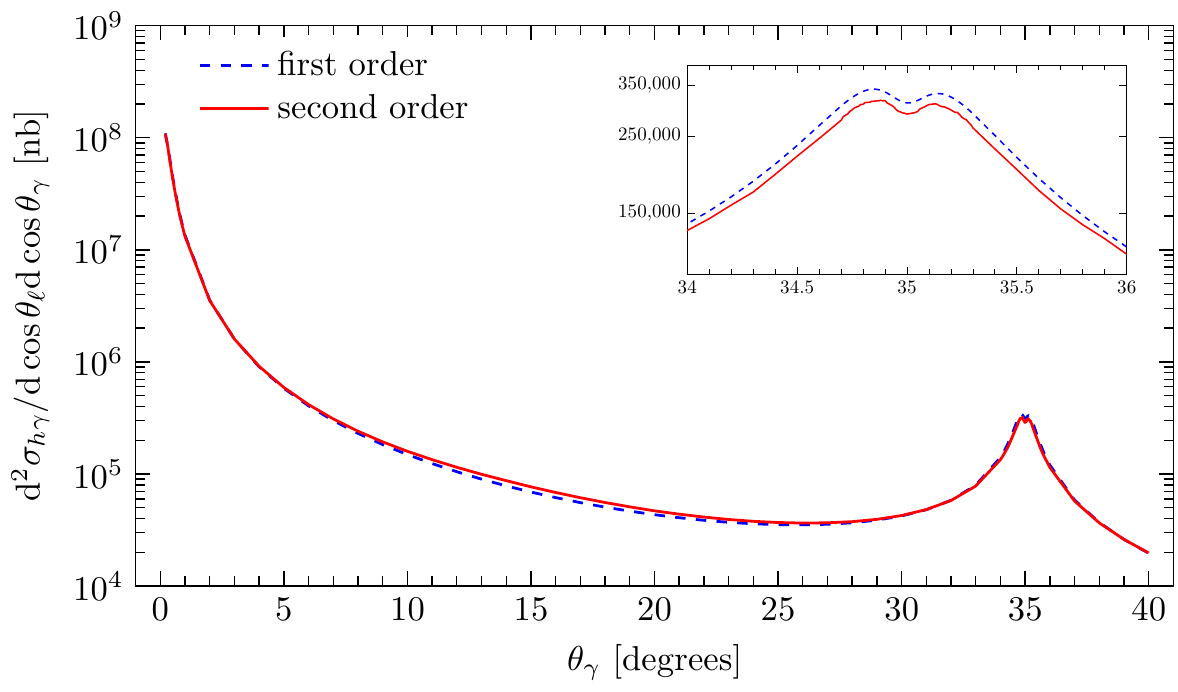}
\caption{
Cross section for radiative scattering with an observed 
photon of energy $E_\gamma \geq \Delta = 10$~MeV at tree 
level (first order, blue dashed line) and including the 
effect due one-loop corrections and a second unobserved 
soft photon with energy below the cut-off $\Delta = 10$~MeV 
(second order, full red line). The beam energy is $E = 
155$~MeV and the scattering angle of the electron fixed 
at $\theta_\ell = 35^\circ$. The energy of the scattered 
electron is integrated over the range $E' > 45$~MeV. 
} 
\label{fig_delta_rad}
\end{figure}

For a better understanding and for completeness we 
also show results for the cross section of radiative 
ep scattering in Fig.~\ref{fig_delta_rad}. This process 
is, in fact, not measurable at P2. The plot of this 
figure shows that bremsstrahlung is dominated by far 
by the emission of photons collinear with the incoming 
electrons, but there is also a peak in the angular 
distribution where photons are emitted in the direction 
of the scattered electron. The one-loop and soft-photon 
corrections for radiative scattering are negative on the 
collinear peaks ($-7\,\%$ for $\theta_\gamma = \theta_\ell 
= 35^\circ$) and positive for photon emission angles far 
away from the peaks ($+11.4\,\%$ for $\theta_\gamma \simeq 
15^\circ$). A particularly interesting feature is a dip on 
top of the final-state radiation peak, shown in more detail 
in the upper right corner of Fig.~\ref{fig_delta_rad}. The 
final-state peak is determined by terms proportional to 
$1/(l^\prime k)$ in the soft-photon eikonal factor, see 
Eq.\ (\ref{eq_cs_soft_photon}), 
\begin{equation}
\frac{1}{l^\prime k} = 
\frac{1}{E_\gamma \left(E' - |\vec{l}^{\, \prime}|\cos \psi 
\right)} \, ,
\end{equation}
where $\psi$ is the angle between the scattered lepton and 
the emitted photon. It is obvious that for a photon emitted 
collinearly with the final-state electron, i.e.\ for $\psi 
\rightarrow 0$, and for $m_\ell \rightarrow 0$ this term 
diverges. One can show that for a finite value of the 
electron mass the dominating terms in the eikonal factor 
vanish for zero emission angle, 
\begin{equation}
\frac{2 l^\prime l}{(l^\prime k)(lk)} 
- \frac{m_\ell^2}{(l^\prime k)^2}
\longrightarrow 
\frac{1}{E_\gamma^2} \frac{4E^{\prime\, 4}}{m_\ell^4} 
\psi^2 
\end{equation}
for $\psi \ll m_\ell/E^\prime \ll 1$, which explains 
the local minimum in the angular distribution of 
Fig.~\ref{fig_delta_rad}. The details of this feature 
depend on the value of the lepton mass and will be 
particularly important for muon scattering. Effects 
due to lepton-mass dependent terms in the cross section 
have been discussed in detail also in 
Refs.~\cite{Barbaro:2013waa,Talukdar:2017dhw}. Our results 
agree with these references. 

\begin{figure}[t]
\centering
\includegraphics[width=0.45\textwidth]{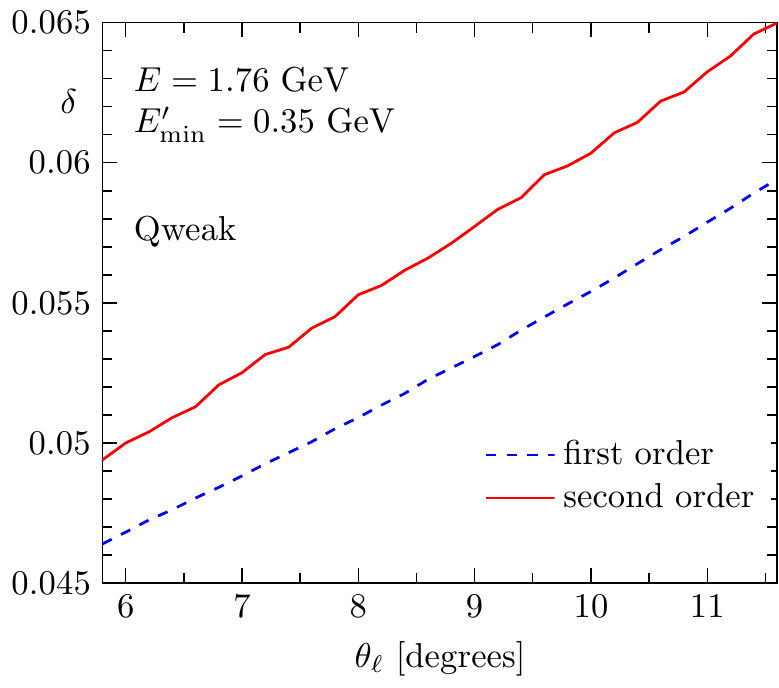}	
\caption{
The complete radiative correction factors at first 
(dashed, blue curve) and second order (full, red curve) 
in the range of scattering angles relevant for Qweak 
at $E = 1.76$~GeV. Hard-photon radiation is included 
with the restriction $E^\prime_{\rm min} = 0.35$~GeV. 
}
\label{fig_delta_tot_Qweak}
\end{figure}

In Figure~\ref{fig_delta_tot_Qweak} we show an example 
of results for the radiative correction factor relevant 
for the Qweak experiment. Here the beam energy is 
$E = 1.16$~GeV and the experiment covers scattering 
angles between $5.8^\circ$ and $11.6^\circ$. The 
corrections are similar to the case of P2 and reach the 
level of roughly 5\,\%. The corrections at second order 
are smaller than at first order by an order of magnitude. 

\begin{figure}[b!]
\centering
\includegraphics[width=0.5\textwidth]{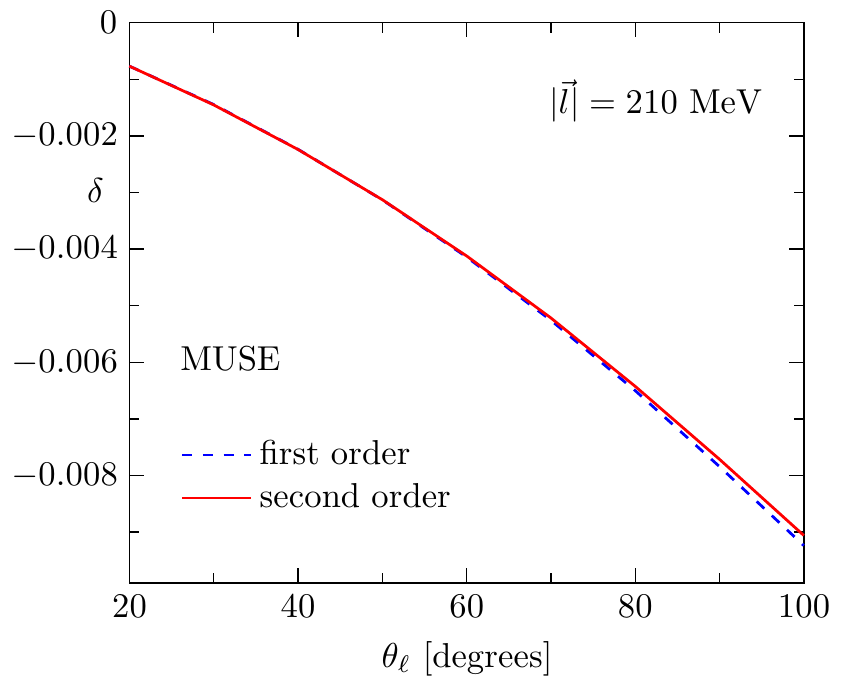}
\caption{
The complete radiative correction factors at first 
(dashed, blue curve) and second order (full, red curve) 
in the range of scattering angles relevant for muon 
scattering at the planned MUSE experiment. Hard-photon 
radiation is included without an additional restriction, 
i.e.\ $E^\prime_{\rm min} = m_\mu$. The soft-photon 
cut-off is fixed at $\Delta = 0.01$~MeV and we have 
used Eq.~(\ref{Eq:delta-2loop}) for the 2-loop 
corrections \cite{Hill:2016gdf}. 
}
\label{fig_delta_tot_MUSE}
\end{figure}

\begin{figure}[t]
\centering
\includegraphics[width=0.49\textwidth]{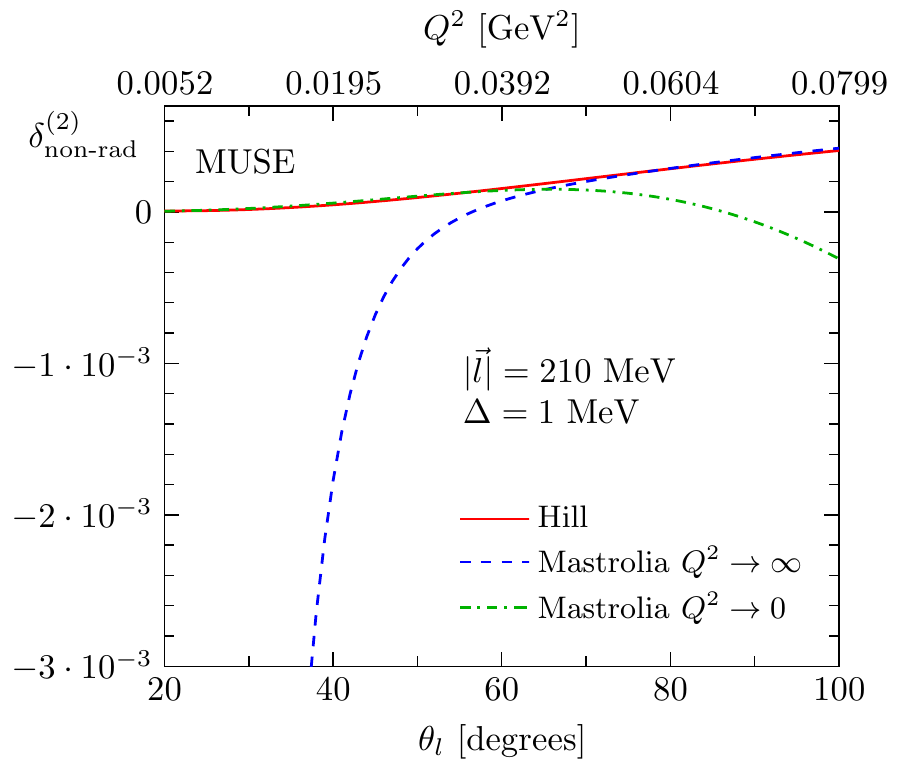}	
\includegraphics[width=0.49\textwidth]{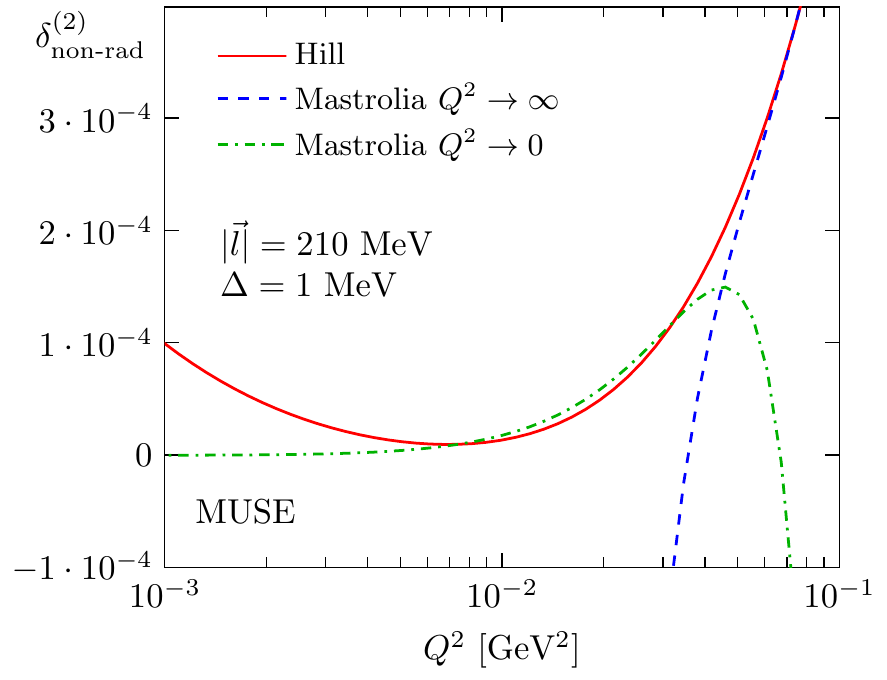}
\caption{
The second-order non-radiative part of the corrections 
for muon scattering at the MUSE experiment using 
different approximations from 
Refs.~\cite{Mastrolia:2003yz,Hill:2016gdf} as described 
in the text. 
}
\label{fig_delta_nonrad1_MUSE}
\end{figure}

Finally we conclude the discussion with a few results 
for the planned MUSE experiment where also the scattering 
of muons off protons will be measured. The beam momentum 
is fixed at $|\vec{l}| = 210$~MeV and we do not impose 
a restriction on the energy of the final-state muon. 
Results are shown in Fig.~\ref{fig_delta_tot_MUSE}. 
The corrections are rather small and vary between 
$-0.1\,\%$ and $-0.9\,\%$. For the calculation of the 
second-order corrections in this figure we have used 
the expression Eq.~(\ref{Eq:delta-2loop}) taken from 
Ref.~\cite{Hill:2016gdf}. This formula is known to be 
valid for large momentum transfer and maybe not applicable 
in the range of scattering angles in the MUSE experiment. 
As a test we show in Fig.~\ref{fig_delta_nonrad1_MUSE} 
the non-radiative correction factor over a larger range 
of $Q^2$ values and using two alternative expressions 
taken from Ref.~\cite{Mastrolia:2003yz}. The one denoted 
``Mastrolia $Q^2\rightarrow \infty$'' contains additional 
lepton mass dependent terms up to and including 4th powers 
of $\mu = \left(m_\ell^2/Q^2\right)$ and the option denoted 
``Mastrolia $Q^2\rightarrow 0$'' includes terms up to 
$\mu^{-4}$. Our expression derived from Hill's result, 
Eq.~(\ref{Eq:delta-2loop}), agrees with 
Ref.~\cite{Mastrolia:2003yz} at large $Q^2$, as expected, 
and seems to provide a nice interpolation between the 
large-$Q^2$ and small-$Q^2$ limits of Ref.~\cite{Mastrolia:2003yz}, 
but this may be accidental. For a conclusive interpretation 
of a high-precision measurement of muon scattering, a 
calculation of 2-loop and 2-photon corrections taking into 
account the full mass dependence will be needed. We mention 
that radiative corrections for the MUSE experiment based 
on an alternative approach have also been studied recently  
in Ref.~\cite{Talukdar:2018hia}.  


\section{Concluding remarks}
\label{sec_conclusion}

We have calculated leptonic QED corrections for elastic 
lepton nucleon scattering at first and second order, i.e.\ 
including one- and two-loop virtual corrections and one- 
and two-photon real radiation. Our study of numerical 
results, in particular for the P2 experiment in Mainz, 
shows that radiative corrections have to be analysed 
with care taking into account experimental details which 
restrict the phase space for photon radiation. Second-order 
corrections are generally small and will become relevant 
for measurements at the per mill level. We plan to make 
the Monte Carlo simulation program, which implements these 
corrections, publicly available in the near future. 

At the time being, our calculation is restricted to the 
leptonic part of corrections. The inclusion of radiation 
at the nucleon and two-photon exchange processes will 
require additional work. In particular, a well-defined 
separation of corrections contained in the effective 
nucleon form factors from corrections which can be 
subtracted during data analysis has still to be worked 
out. The extension of the calculation for the polarization 
dependent part, i.e.\ including Feynman diagrams with 
$Z$-boson exchange, is in preparation. This calculation 
will be needed for the planned high-precision measurement 
of the weak charge of the proton at the P2 experiment.

\clearpage

\begin{appendices}

\section{Cross-sections for $ep\rightarrow ep$ and 
$ep\rightarrow ep\gamma$}
\label{app_me}

The matrix element squared for non-radiative ep scattering, 
averaged and summed over the spin degrees of freedom in the 
initial and final states, can be given in a compact form 
using the following form factor combinations: 
\begin{equation}
G \equiv - 2 Q^2 \left(F_1^p + F_2^p\right)^2 \, , 
\quad \quad 
H \equiv 4\left(F_1^p\right)^2 
+ \frac{Q^2}{M^2} \left(F_2^p\right)^2 \, .
\end{equation}
We find 
\begin{equation}
\frac{\mathrm{d}\sigma^{(0)}}{\mathrm{d}Q^2} 
= 
\frac{\overline{|\mathcal{M}_\text{Born}|^2}}{16\pi 
\left[(s - m_\ell^2 - M^2)^2 - m_\ell^2 M^2 \right]}  
\end{equation}
with
\begin{equation}
\overline{\left|\mathcal{M}_\text{Born}\right|^2} 
=
\frac{(4\pi\alpha)^2}{Q^4} 
\left\{ 
G(2m_e^2-Q^2)
- H \left[M^2Q^2 
+ \left(s-m_e^2-M^2\right)\left(Q^2-s+m_e^2+M^2\right)\right] 
\right\} \, ,
\end{equation}
and $s = (l+p)^2$. For the matrix element squared of the 
radiative process with one additional photon, it is 
convenient to introduce also the variables $S = 2 l_1 
\cdot p_1$ and $U = - 2 l_2 \cdot p_1$. We find for the 
matrix element squared for $ep \rightarrow ep\gamma$, 
averaged and summed over initial and final state spin 
degrees of freedom:
\begin{eqnarray}
\overline{\left|\mathcal{M}_{1\gamma}\right|^2} 
&=&
\frac{(4\pi\alpha)^3}{Q^4} 
\frac{1}{(l_1k)^2(2l_1k+Q^2_\ell-Q^2)^2}
\bigg\{ 
\\ && \quad
G \Big[- 16(l_1k)^4 - 16(l_1k)^3(Q^2_\ell - Q^2)
+ (l_1k)^2(8m_\ell^2Q^2_\ell - 6Q^4_\ell + 8Q^2_\ell Q^2 - 6Q^4) 
\nonumber \\ && \quad \quad
- l_1k(Q^2_\ell - Q^2)(-4m_\ell^2Q^2_\ell + Q^4_\ell + Q^4) 
- m_\ell^2(2m_\ell^2 - Q^2)(Q^2_\ell - Q^2)^2\Big] 
\nonumber \\ && \quad
- H \Big[
16(l_1k)^4M^2 + 8(l_1k)^3 \big(2M^2(Q^2_\ell - Q^2) 
+ \frac{1}{2} Q^2 (S - U)
\big)
\nonumber \\ && \quad \quad
+ 2(l_1k)^2 \left(2m_\ell^2(Q^2 - S - U)^2 
+ Q^2Q^2_\ell(2S - U) + Q^4(S - 2U) \right. 
\nonumber \\ && \quad \quad \left.
+ Q^2(S^2 + U^2) + M^2(3Q^4_\ell - 4Q^2_\ell Q^2 + 3Q^4)
\right) 
\nonumber \\ && \quad \quad
+ l_1k (Q^2_\ell - Q^2) \left(2m_\ell^2(Q^2 - 2U)(Q^2 - S - U)
\right.
\nonumber \\ && \quad \quad \left.
- Q^2(- Q^2_\ell S - Q^2 U + S^2 + U^2) 
+ M^2(Q^4_\ell + Q^4) \right) 
\nonumber \\ && \quad \quad
- m_\ell^2(Q^2_\ell - Q^2)^2
\left(M^2Q^2 + U(Q^2 - U)\right) \Big] \bigg\} \, .
\nonumber
\end{eqnarray}

\section{Phase-space for one-photon bremsstrahlung, 
$ep\rightarrow ep\gamma$}
\label{app_1g-phase-space}

After integrating out the final-state nucleon momentum to 
remove the $\delta$-function for energy-momentum conservation, 
the phase space for $\ell p\rightarrow \ell p\gamma$ is given 
by
\begin{equation}
\mathrm{d}^4\Gamma 
= 
\int \frac{1}{(2 \pi)^5} 
\frac{\mathrm{d}^3 l^\prime}{2 E'} 
\frac{\mathrm{d}^3 k}{2 E_\gamma} 
\delta \left[ (l + p - l^\prime - k)^2 - M^2 \right] 
\, . 
\end{equation}
Using energies and angles as shown in Fig.~\ref{Fig:process_kin}
we write 
\begin{align}
\nonumber
&\mathrm{d}^3 l^\prime 
= E' \, |\vec{l}^{\, \prime}| \, 
\mathrm{d} \cos \theta_\ell \, 
\mathrm{d} \phi_\ell \, 
\mathrm{d} E' \, , 
\\
\nonumber
&\mathrm{d}^3 k 
= E_\gamma^2 \, 
\mathrm{d} \cos \theta_\gamma \, 
\mathrm{d} \phi_\gamma \, 
\mathrm{d} E_\gamma \, , 
\\
&(l + p - l^\prime - k)^2 - M^2 = 2 (A - B \cos \phi_\gamma) 
\end{align}
with 
\begin{align}
\nonumber
A = & 
M (E-E'-E_\gamma) - EE' + E'E_\gamma - EE_\gamma 
+ |\vec{l}||\vec{l}^{\, \prime}| \cos \theta_\ell 
\\
\nonumber 
&+ |\vec{l}|E_\gamma \cos \theta_\gamma 
- E_\gamma |\vec{l}^{\, \prime}| \cos \theta_\ell \cos \theta_\gamma 
+ m_\ell^2 \, , 
\\
B = & 
E_\gamma |\vec{l}^{\, \prime}| \sin \theta_\ell \sin \theta_\gamma 
\, .
\label{Eq:AandB}
\end{align}
For unpolarized scattering, one can perform the integration 
over the azimuthal angle of the scattered lepton $\phi_\ell$ 
and we find 
\begin{align}
\nonumber
\mathrm{d}\Gamma 
&= 
\int \frac{1}{4(2 \pi)^4} 
E_\gamma |\vec{l}^{\, \prime}| \, 
\mathrm{d} E' \, 
\mathrm{d} \cos \theta_\ell \, 
\mathrm{d} E_\gamma \, 
\mathrm{d} \cos \theta_\gamma \, 
\mathrm{d} \phi_\gamma \, 
\delta \left[ 2(A-B \cos \phi_\gamma) \right] 
\\
&= 
\frac{1}{8(2 \pi)^4} \, 
\frac{\mathrm{d} E' \, 
\mathrm{d} \cos \theta_\ell \, 
\mathrm{d} E_\gamma \, 
\mathrm{d} \cos \theta_\gamma}%
{\sin \theta_\ell \sin \theta_\gamma \sin \phi_\gamma} \,  
\Theta \left(1 - \frac{A^2}{B^2} \right) 
\, ,
\end{align} 
where $\sin \phi_\gamma = \sqrt{1 - A^2/B^2} $.
Integration limits follow from the condition 
\begin{equation}
\frac{A^2}{B^2} \leq 1 \, . 
\end{equation}
We find 
\begin{align}
\label{Eq:1hg-limits}
& 
m_\ell < E' < E \, ,
\\
\nonumber
&
\max \left( 
\frac{EE' - M(E-E') - m_\ell^2}{|\vec{l}||\vec{l}^{\, \prime}|} 
\, , 
-1 \right) 
< 
\cos \theta_\ell 
< 1 \, , 
\\[1ex]
&
\frac{y}{a-x} < E_\gamma < -\frac{y}{a+x} \, , 
\nonumber
\\[1ex]
&
\frac{-F\sqrt{F^2+D^2-C^2}-DC}{D^2+F^2} 
< 
\cos \theta_\gamma 
< \frac{F\sqrt{F^2+D^2-C^2}-DC}{D^2+F^2} \, , 
\nonumber 
\end{align}
where we have used 
\begin{align*}
C =& 
M (E-E'-E_\gamma) - EE' + E'E_\gamma - EE_\gamma 
+ |\vec{l}||\vec{l}^{\, \prime}|\cos \theta_\ell  + m_\ell^2 \, , 
\\
D =& 
E_\gamma (|\vec{l}| - |\vec{l}^{\, \prime}| \cos \theta_\ell) \, , 
\\
F =& 
E_\gamma |\vec{l}^{\, \prime}| \sin \theta_\ell \, , 
\\
a =& 
\sqrt{|\vec{l}|^2 + |\vec{l}^{\, \prime}|^2 
- 2|\vec{l}||\vec{l}^{\, \prime}|\cos \theta_\ell} \, , 
\\
x =& 
E'-E-M \, , 
\\
y =& 
M(E-E') - EE' + |\vec{l}||\vec{l}^{\, \prime}| \cos \theta_\ell 
+ m_\ell^2 \, .
\end{align*}

\section{Phase-space for two-photon bremsstrahlung, 
$ep\rightarrow ep\gamma\gamma$}
\label{app_2g-phase-space}

We use the same notation as shown in Fig.~\ref{Fig:process_kin}, 
but consider two photons in the final state, whose 
4-momenta are denoted by $k$ and $k^\prime$ with energies 
$E_{\gamma}$, $E_{\gamma}^\prime$ and angles $\theta_{\gamma}$, 
$\theta_{\gamma}^\prime$, $\phi_{\gamma}$, and 
$\phi_{\gamma}^\prime$, respectively. Using the $\delta$ function 
from energy-momentum conservation, the integration over the 
4-particle phase space can be written as 
\begin{equation}
\mathrm{d}^7\Gamma_{2\gamma} 
= 
\frac{E_\gamma E'_\gamma |\vec{l}^{\, \prime}|}{16(2 \pi)^7}  
\frac{\mathrm{d} E' \, 
\mathrm{d} \cos \theta_\ell \, 
\mathrm{d} E_\gamma \, 
\mathrm{d} E'_\gamma \, 
\mathrm{d} \cos \theta'_\gamma \, 
\mathrm{d} \phi'_\gamma \, 
\mathrm{d} \cos \theta_\gamma }%
{\left|\alpha_1 \cos \phi_\gamma - \alpha_2 \sin \phi_\gamma \right|} 
\end{equation}
where 
\begin{equation}
\label{phi_gamma}
\phi_\gamma 
= 
\arcsin \frac{\alpha_3}{\sqrt{\alpha^2_1 + \alpha^2_2}} 
- \arcsin \frac{\alpha_2}{\sqrt{\alpha^2_1 + \alpha^2_2}} 
\end{equation}
with
\begin{align*}
\alpha_1 
= & 
E_\gamma E'_\gamma \sin\theta_\gamma \sin\theta'_\gamma 
\sin\phi'_\gamma \, , 
\\
\alpha_2 
= & 
|\vec{l}^{\, \prime}| E_\gamma \sin\theta_\ell \sin\theta_\gamma 
+ E_\gamma E'_\gamma \sin\theta_\gamma \sin \theta'_\gamma 
\cos \phi'_\gamma \, , 
\\
\alpha_3 
= & 
m_\ell^2 + M(E - E' - E_\gamma - E'_\gamma) 
- E(E' + E_\gamma + E'_\gamma) + E'(E_\gamma + E'_\gamma) 
+ |\vec{l}||\vec{l}^{\, \prime}| \cos \theta_\ell 
\\
&+ |\vec{l}| E_\gamma \cos\theta_\gamma 
+ |\vec{l}| E'_\gamma \cos\theta'_\gamma 
- |\vec{l}^{\, \prime}| E'_\gamma (\sin\theta_\ell \sin\theta'_\gamma 
\cos\phi'_\gamma + \cos\theta_\ell \cos\theta'_\gamma) 
\\
& - |\vec{l}^{\, \prime}|E_\gamma \cos\theta_\ell \cos\theta_\gamma 
- E_\gamma E'_\gamma \cos\theta_\gamma \cos\theta'_\gamma 
+ E_\gamma E'_\gamma 
\, .
\end{align*}
Integration limits for angles and energies follow from 
the condition that the arguments of the $\arcsin$-functions 
in the expression for $\phi_\gamma$ in Eq.~(\ref{phi_gamma}) 
have to be in the allowed range between $-1$ and $+1$. 
The required calculations are straightforward, but tedious, 
and we write down only a few partial results in the following. 
For $\theta_\gamma$ we find:  
\begin{equation} 
\frac{- \beta_2 \beta_3 - \beta_1 \sqrt{\beta}}%
{\beta_1^2 + \beta_2^2} 
< 
\cos\theta_\gamma 
< 
\frac{- \beta_2 \beta_3 + \beta_1 \sqrt{\beta}}%
{\beta_1^2 + \beta_2^2} 
\end{equation}
with 
\begin{align*}
\beta_1 
= & 
\sqrt{2 E_\gamma^2 |\vec{l}^{\, \prime}| E'_\gamma 
\sin\theta_\ell \sin\theta'_\gamma \cos\phi'_\gamma 
+ E_\gamma^2 E_\gamma^{\prime \, 2} \sin^2\theta'_\gamma 
+ |\vec{l}^{\, \prime}|^2 E_\gamma^2 \sin^2\theta_\ell} \, ,
\\
\beta_2 
= & 
|\vec{l}| E_\gamma - |\vec{l}^{\, \prime}| E_\gamma 
\cos\theta_\ell - E_\gamma E'_\gamma \cos\theta'_\gamma \, ,
\\
\beta_3 
= & 
m_\ell^2 + M(E - E' - E_\gamma - E'_\gamma) 
- E(E' + E_\gamma + E'_\gamma) + E'(E_\gamma + E'_\gamma) 
+ |\vec{l}||\vec{l}^{\, \prime}| \cos \theta_\ell 
\\
&+ |\vec{l}| E'_\gamma \cos\theta'_\gamma 
- |\vec{l}^{\, \prime}| E'_\gamma (\sin\theta_\ell \sin\theta'_\gamma 
\cos\phi'_\gamma + \cos\theta_\ell \cos\theta'_\gamma) 
+ E_\gamma E'_\gamma \, .
\end{align*}
The integration limits for $\phi'_\gamma$ are: 
\begin{equation}
\frac{E_\gamma^2 + \gamma_2 - E_\gamma \sqrt{\gamma}}%
{\gamma_1} 
< \cos\phi'_\gamma < 
\frac{E_\gamma^2 + \gamma_2 + E_\gamma \sqrt{\gamma}}%
{\gamma_1}  
\end{equation}
with 
\begin{equation}
\gamma = 2\gamma_2 + \gamma_3 > 0 
\end{equation}
and 
\begin{align}
\gamma_1 
= &
|\vec{l}^{\, \prime}| E'_\gamma \sin\theta_\ell \sin\theta'_\gamma 
\, , 
\\
\gamma_2
= & m_\ell^2 + M(E - E' - E_\gamma - E'_\gamma) 
- E(E' + E_\gamma + E'_\gamma) + E'(E_\gamma + E'_\gamma) 
\\
&+ E_\gamma E'_\gamma 
+ |\vec{l}||\vec{l}^{\, \prime}| \cos \theta_\ell 
+ E'_\gamma \cos\theta'_\gamma 
\left(|\vec{l}| - |\vec{l}^{\, \prime}|\cos \theta_\ell\right) \, , 
\nonumber 
\\
\gamma_3
= &
|\vec{l}|^2 + |\vec{l}^{\, \prime}|^2 + E_\gamma^2 
+ E_\gamma^{\prime \, 2} - 2E'_\gamma \cos\theta'_\gamma 
\left(|\vec{l}| - |\vec{l}^{\, \prime}|\cos \theta_\ell\right) 
- 2|\vec{l}||\vec{l}^{\, \prime}|\cos\theta_\ell 
\, .
\end{align}
The condition $\gamma \geq 0$ is always fulfilled provided 
that 
\begin{equation}
E'_\gamma \leq E - E' - E_\gamma \, .
\end{equation}
In our implementation of the numerical integration over 
the phase space we make sure that no kinematic limit is 
missed by reconstructing always complete events and checking 
the 4-momentum conservation. This allows us to use the phase 
space integrator as an event generator which can be used for 
the simulation of an experiment. It turns out that an explicit 
implementation of the kinematic limits given above in the 
numerical integration routine is sufficient to render the 
efficiency of the Monte Carlo integration at a high level 
above $70\,\%$.


\end{appendices}


\section*{Acknowledgment}

We thank our experimental colleagues of the P2 collaboration 
and in particular D.~Becker for his feedback and help to 
develop a Monte Carlo simulation program suited for usage 
in the experimental environment.  
This work was supported by the Deutsche Forschungsgemeinschaft 
(DFG) in the framework of the collaborative research center 
SFB1044 ``The Low-Energy Frontier of the Standard Model: From 
Quarks and Gluons to Hadrons and Nuclei''.



\begin{thebibliography}{99}

\bibitem{Androic:2018kni}
  D.~Androi\'c {\it et al.} [Qweak Collaboration],
  Nature {\bf 557} (2018) 207.

\bibitem{Becker:2018ggl}
  D.~Becker {\it et al.},
  arXiv:1802.04759 [nucl-ex].

\bibitem{Gilman:2013eiv}
  R.~Gilman {\it et al.} [MUSE Collaboration],
  arXiv:1303.2160 [nucl-ex].

\bibitem{Mo:1968cg}
  L.~W.~Mo and Y.~S.~Tsai,
  Rev.\ Mod.\ Phys.\ {\bf 41} (1969) 205.

\bibitem{Tsai:1961zz}
  Y.~S.~Tsai,
  Phys.\ Rev.\  {\bf 122} (1961) 1898.

\bibitem{Tsai:1971qi}
  Y.~S.~Tsai,
  SLAC-PUB-0848.

\bibitem{deCalan:1990eb}
  C.~de Calan, H.~Navelet and J.~Picard,
  Nucl.\ Phys.\ B {\bf 348} (1991) 47.

\bibitem{Maximon:2000hm}
  L.~C.~Maximon and J.~A.~Tjon,
  Phys.\ Rev.\ C {\bf 62} (2000) 054320
  [nucl-th/0002058].

\bibitem{Weissbach:2004ij}
  F.~Weissbach, K.~Hencken, D.~Rohe, I.~Sick and D.~Trautmann,
  Eur.\ Phys.\ J.\ A {\bf 30} (2006) 477
  [nucl-th/0411033].

\bibitem{Weissbach:2008nx}
  F.~Weissbach, K.~Hencken, D.~Rohe and D.~Trautmann,
  Phys.\ Rev.\ C {\bf 80} (2009) 024602
  [arXiv:0805.1535 [nucl-th]].

\bibitem{Akushevich:2015toa}
  I.~Akushevich, H.~Gao, A.~Ilyichev and M.~Meziane,
  Eur.\ Phys.\ J.\ A {\bf 51} (2015) 1.

\bibitem{Akushevich:2012tw}
  I.~Akushevich and A.~Ilyichev,
  Phys.\ Rev.\ D {\bf 85} (2012) 053008
  [arXiv:1201.4065 [hep-ph]].

\bibitem{Akushevich:2017kct}
  I.~Akushevich and A.~Ilyichev,
  arXiv:1712.00091 [hep-ph].

\bibitem{Arbuzov:2015vba}
  A.~B.~Arbuzov and T.~V.~Kopylova,
  Eur.\ Phys.\ J.\ C {\bf 75} (2015) 603
  [arXiv:1510.06497 [hep-ph]].

\bibitem{Kuraev:2013dra}
  E.~A.~Kuraev, A.~I.~Ahmadov, Y.~M.~Bystritskiy and 
  E.~Tomasi-Gustafsson,
  Phys.\ Rev.\ C {\bf 89} (2014) 065207
  [arXiv:1311.0370 [hep-ph]].

\bibitem{Borisyuk:2014ssa}
  D.~Borisyuk and A.~Kobushkin,
  Phys.\ Rev.\ C {\bf 90} (2014) 025209
  [arXiv:1405.2467 [hep-ph]].

\bibitem{Gerasimov:2016zfr}
  R.~E.~Gerasimov and V.~S.~Fadin,
  J.\ Phys.\ G {\bf 43} (2016) 125003
  [arXiv:1604.07960 [nucl-th]].

\bibitem{Arrington:2011dn}
  J.~Arrington, P.~G.~Blunden and W.~Melnitchouk,
  Prog.\ Part.\ Nucl.\ Phys.\  {\bf 66} (2011) 782
  [arXiv:1105.0951 [nucl-th]].

\bibitem{Ent:2001hm}
  R.~Ent, B.~W.~Filippone, N.~C.~R.~Makins, R.~G.~Milner, 
  T.~G.~O'Neill and D.~A.~Wasson,
  Phys.\ Rev.\ C {\bf 64} (2001) 054610.

\bibitem{Gakh:2016xby}
  G.~I.~Gakh, M.~I.~Konchatnij, N.~P.~Merenkov and 
  E.~Tomasi-Gustafsson,
  Phys.\ Rev.\ C {\bf 95} (2017) 055207
  [arXiv:1612.02139 [hep-ph]].

\bibitem{Kwiatkowski:1990es}
  A.~Kwiatkowski, H.~Spiesberger and H.~J.~M\"ohring,
  Comput.\ Phys.\ Commun.\  {\bf 69} (1992) 155.

\bibitem{Arbuzov:1995id}
  A.~Arbuzov, D.~Y.~Bardin, J.~Bl\"umlein, L.~Kalinovskaya and 
  T.~Riemann,
  Comput.\ Phys.\ Commun.\  {\bf 94} (1996) 128
  [hep-ph/9511434].

\bibitem{Afanasev:2001nn}
  A.~V.~Afanasev, I.~Akushevich, A.~Ilyichev and N.~P.~Merenkov,
  Phys.\ Lett.\ B {\bf 514} (2001) 269
  [hep-ph/0105328].

\bibitem{Charchula:1994kf}
  K.~Charchula, G.~A.~Schuler and H.~Spiesberger,
  Comput.\ Phys.\ Commun.\  {\bf 81} (1994) 381.

\bibitem{Akushevich:2011zy}
  I.~Akushevich, O.~F.~Filoti, A.~N.~Ilyichev and N.~Shumeiko,
  Comput.\ Phys.\ Commun.\  {\bf 183} (2012) 1448
  [arXiv:1104.0039 [hep-ph]].

\bibitem{anaximandros} 
  R.~D.~Bucoveanu and H.~Spiesberger, in preparation. 

\bibitem{Drell:1963ej}
  S.~D.~Drell and J.~D.~Walecka,
  Annals Phys.\  {\bf 28} (1964) 18.

\bibitem{Shtabovenko:2016sxi}
  V.~Shtabovenko, R.~Mertig and F.~Orellana,
  Comput.\ Phys.\ Commun.\  {\bf 207} (2016) 432
  [arXiv:1601.01167 [hep-ph]].

\bibitem{Hahn:1998yk}
  T.~Hahn and M.~Perez-Victoria,
  Comput.\ Phys.\ Commun.\  {\bf 118} (1999) 153
  [hep-ph/9807565].

\bibitem{Vanderhaeghen:2000ws}
  M.~Vanderhaeghen, J.~M.~Friedrich, D.~Lhuillier, D.~Marchand, 
  L.~Van Hoorebeke and J.~Van de Wiele,
  Phys.\ Rev.\ C {\bf 62} (2000) 025501
  [hep-ph/0001100].

\bibitem{Bytev:2003qf}
  V.~V.~Bytev, E.~A.~Kuraev and E.~Tomasi-Gustafsson,
  Phys.\ Rev.\ C {\bf 77} (2008) 055205
  [hep-ph/0310226].

\bibitem{tHooft:1978jhc}
  G.~'t Hooft and M.~J.~G.~Veltman,
  Nucl.\ Phys.\ B {\bf 153} (1979) 365.

\bibitem{Gramolin:2014pva}
  A.~V.~Gramolin, V.~S.~Fadin, A.~L.~Feldman, R.~E.~Gerasimov, 
  D.~M.~Nikolenko, I.~A.~Rachek and D.~K.~Toporkov,
  J.\ Phys.\ G {\bf 41} (2014) 115001
  [arXiv:1401.2959 [nucl-ex]].

\bibitem{Hahn:2004fe}
  T.~Hahn,
  Comput.\ Phys.\ Commun.\  {\bf 168} (2005) 78
  [hep-ph/0404043].

\bibitem{Kallen:1955fb}
  A.~O.~G.~Kallen and A.~Sabry,
  Kong.\ Dan.\ Vid.\ Sel.\ Mat.\ Fys.\ Med.\ {\bf 29} (1955) 17, 1.

\bibitem{ignatov-vpl}
  F.~Ignatov, 
  %
  \url{http://cmd.inp.nsk.su/~ignatov/vpl/}

\bibitem{Actis:2010gg}
  S.~Actis {\it et al.} [Working Group on Radiative Corrections 
  and Monte Carlo Generators for Low Energies],
  Eur.\ Phys.\ J.\ C {\bf 66} (2010) 585
  [arXiv:0912.0749 [hep-ph]].

\bibitem{Jegerlehner:2011mw}
  F.~Jegerlehner,
  Nuovo Cim.\ C {\bf 034S1} (2011) 31
  [arXiv:1107.4683 [hep-ph]].

\bibitem{Keshavarzi:2018mgv}
  A.~Keshavarzi, D.~Nomura and T.~Teubner,
  Phys.\ Rev.\ D {\bf 97} (2018) no.11,  114025
  [arXiv:1802.02995 [hep-ph]].

\bibitem{Hill:2016gdf}
  R.~J.~Hill,
  Phys.\ Rev.\ D {\bf 95} (2017) 013001
  [arXiv:1605.02613 [hep-ph]].

\bibitem{vanRitbergen:1998hn}
  T.~van Ritbergen and R.~G.~Stuart,
  Phys.\ Lett.\ B {\bf 437} (1998) 201
  [hep-ph/9802341].

\bibitem{Davydychev:2000ee}
  A.~I.~Davydychev, K.~Schilcher and H.~Spiesberger,
  Eur.\ Phys.\ J.\ C {\bf 19} (2001) 99
  [hep-ph/0011221].

\bibitem{Mastrolia:2003yz}
  P.~Mastrolia and E.~Remiddi,
  Nucl.\ Phys.\ B {\bf 664} (2003) 341
  [hep-ph/0302162].

\bibitem{Bonciani:2003ai}
  R.~Bonciani, P.~Mastrolia and E.~Remiddi,
  Nucl.\ Phys.\ B {\bf 676} (2004) 399
  [hep-ph/0307295].

\bibitem{Barbaro:2013waa}
  M.~B.~Barbaro, C.~Maieron and E.~Voutier,
  Phys.\ Lett.\ B {\bf 726} (2013) 505, 
  Erratum: [Phys.\ Lett.\ B {\bf 727} (2013) 573]
  [arXiv:1305.3873 [hep-ph]].

\bibitem{Talukdar:2017dhw}
  P.~Talukdar, F.~Myhrer and U.~Raha,
  arXiv:1712.09963 [nucl-th].

\bibitem{Talukdar:2018hia}
  P.~Talukdar, F.~Myhrer, G.~Meher and U.~Raha,
  arXiv:1810.04027 [nucl-th].


\end{thebibliography}

\end{document}